\documentclass{article}
\usepackage{slashed,verbatim,subfigure}
\usepackage[numbers,sort&compress]{natbib}
\usepackage{amsmath}
\usepackage{amssymb}
\usepackage{appendix}
\usepackage{graphicx}
\usepackage{dcolumn}
\usepackage{bm}
\usepackage[colorlinks=true,linktocpage=true,
linkcolor=blue,citecolor=blue]{hyperref}
\usepackage[a4paper]{geometry}
\newcommand{\be}{\begin{eqnarray}}
\newcommand{\ee}{\end{eqnarray}}
\begin{document}
\large
\title{\bf{Thermomagnetic properties and Bjorken expansion of hot QCD matter 
in a strong magnetic field}}
\author{Shubhalaxmi Rath\footnote{srath@ph.iitr.ac.in}~~and~~Binoy Krishna 
Patra\footnote{binoyfph@iitr.ac.in}\vspace{0.1in}\\ 
Department of Physics,\\ 
Indian Institute of Technology Roorkee, Roorkee 247667, India}
\date{}
\maketitle
\begin{abstract}
In this work we have studied the effects of an external 
strong magnetic field on the thermodynamic and magnetic properties of a
hot QCD matter and then explored these effects on the 
subsequent hydrodynamic expansion of the said matter once 
produced in the ultrarelativistic heavy ion collisions.
For that purpose, we have computed the quark and gluon self-energies 
up to one loop in the strong magnetic field, using the 
HTL approximation with two hard scales - temperature and magnetic field, 
which in turn 
compute the effective propagators for quarks and gluons, respectively. 
Hence the quark and gluon contributions to the free energy are
obtained from the respective propagators and finally derive the 
equation of state (EOS) by calculating the pressure and energy 
density. We have found that the speed of sound is enhanced due to the presence 
of strong magnetic field and this effect
will be later exploited in the hydrodynamics. Thereafter the
magnetic properties are studied from the free energy
of the matter, where the magnetization is found to increase linearly 
with the magnetic field, thus hints the paramagnetic behavior. The 
temperature dependence of the magnetization is also studied
within the strong magnetic field limit ($|q_fB| \gg T^2$, 
$|q_fB| \gg m_f^2$, where $|q_f| (m_f)$ is the absolute electric
charge (mass) of $f$-th flavor), where the magnetization is found to 
increase slowly with the temperature.
Finally, to see how a strong magnetic field could affect the 
hydrodynamic evolution, 
we have revisited the Bjorken boost-invariant picture with our 
paramagnetic EOS as an input in the equation of motion for the energy-momentum 
conservation. We have noticed that the energy density evolves faster than 
in the absence of strong magnetic field, {\em i.e.} cooling becomes faster, 
which could have implications on the heavy-ion phenomenology. As mentioned
earlier, this observation can be understood by the enhancement of the speed of sound.

\end{abstract}

{\bf Keywords:} Self energy; Free energy; Equation of state; Strong magnetic field; Magnetization; Susceptibility; Bjorken expansion.

\newpage

\section{Introduction}
The possibility to achieve strong magnetic fields at the noncentral 
events of ultrarelativistic heavy ion collisions 
(URHIC) \cite{Tuchin:PRC83'2011,Fukushima:PRD86'2012} {\em viz.}
at RHIC, LHC, makes it feasible in reality to explore the thermodynamic 
and magnetic properties of QCD matter in the presence of both heat-bath 
and strong magnetic field. Recently the lattice QCD simulations have 
determined the equation of state (EOS) for a (2+1)-flavor QCD in the 
presence of magnetic fields 
by calculating the thermodynamic observables 
\cite{Bali:JHEP1408'2014,Endrodi:JHEP1507'2015}. 
Similarly the effects 
of strong magnetic fields on the phase structures of the hadronic 
matter in the framework of effective theories have also been 
reviewed in \cite{Andersen:RMP88'2016}.
We have very recently investigated the 
thermodynamic observables for the same perturbatively in a strong 
magnetic field (SMF) limit \cite{Rath:JHEP1712'2017}. 
Like the thermodynamic properties, the magnetic properties of a hot QCD 
matter might be modified due to the presence of strong magnetic 
field. Different materials placed in an external magnetic field 
evince different responses to the magnetic field and depending on
the responses, known as magnetization, they are 
classified as paramagnets and diamagnets. 
Paramagnetic response is defined by the positive value 
of the magnetization, {\em i.e.} the case when the free energy of the system 
decreases due to the magnetic field, whereas the negative value of the 
magnetization defines the system or material as a diamagnet. The 
classification of magnetic materials can be alternatively described in 
terms of magnetic susceptibility. So the magnetic properties of 
QCD matter too can also be 
envisaged by its response to the external magnetic field
and can thus be understood by the free energy of the system, which in 
general embraces all the informations about the magnetic and thermodynamic 
properties of a system. The knowledge about the magnetic response of the QCD 
matter helps greatly in exploring the chiral magnetic effect  \cite{Fukushima:PRD78'2008,Kharzeev:NPA803'2008}, the 
photon production rate from the hot QCD medium 
\cite{Hees:PRC84'2011,Shen:PRC89'2014}, the dilepton production from the QGP \cite{Tuchin:PRC88'2013,Mamo:JHEP1308'2013}, the collective expansion of the 
fireball \cite{Mohapatra:MPLA26'2011}, the axial magnetic 
field \cite{Braguta:PRD89'2014,Chernodub:PRB89'2014}, the chiral vortical 
effect \cite{Kharzeev:PRL106'2011}, the (inverse) magnetic catalysis arising 
due to the (restoration) breaking of the chiral symmetry \cite{Mueller:PRD91'2015,Gusynin:PRL73'1994,Haber:PRD90'2014}, the bulk 
properties of a Fermi gas \cite{Strickland:PRD86'2012}, the 
production of soft photons in heavy ion collisions  \cite{Basar:PRL109'2012,Ayala:EPJWC141'2017}, the Faraday rotation in a 
magnetized thermal medium \cite{D’Olivo:PRD67'2003,Ganguly:PRD60'1999} and the  electromagnetic radiation \cite{Tuchin:PRC87'2013}. The magnetic response of 
QCD matter behaves as a theoretical tool in the area of high energy physics 
where strong  magnetic fields might exist, such as the beginning of the 
universe and the cores of the magnetars, in addition to RHIC and LHC.

Thus we get motivated to analyse the magnetic response of a hot QCD matter
in an ambience of strong magnetic field, which have been studied 
previously in different effective theories, such as in the 
potential model \cite{Kabat:PRD66'2002}, the holographic 
QCD \cite{Bergman:JHEP05'2008}, Fermi-liquid theory by evaluating the spin 
susceptibility \cite{Sato:PTPS174'2008}, the transport model 
\cite{Steinert:PRC89'2014}, the linear sigma model with the Polyakov loop \cite{Tawfik:PRC90'2014} and the quark-meson model applying the 
functional renormalization group method \cite{Kamikado:JHEP01'2015}. 
The above lattice study \cite{Bali:JHEP1408'2014} also investigates the effect 
of background magnetic field on the magnetization from the free energy. However, 
our aim is to find the effects of strong magnetic field on the magnetic properties
of the hot QCD matter as well as on the dynamical evolution of the same matter. 
So we have first revisited the free energy of the aforesaid matter 
in strong magnetic field ($|q_fB| \gg T^2$, $|q_fB| \gg m_f^2$) 
perturbatively up to one loop in HTL approximation with two 
hard scales - one is the temperature for gluons and the other is the 
strong magnetic field for quarks. The free energy further derives 
the EOS, where we have found that the square of the speed of sound 
gets enhanced due to the presence 
of strong magnetic field. This observation will be exploited in 
the expansion of the matter.
We will next study the magnetic properties 
through the magnetization and magnetic
susceptibility, which are obtained from the abovementioned perturbative 
free energy. We have found that the magnetization thus obtained indicates
the paramagnetic nature of the hot QCD matter. 

So far we have discussed the thermodynamic and magnetic 
properties of the medium, assuming the medium as static. But the medium 
produced in the ultrarelativistic heavy ion collisions immediately 
expands hydrodynamically. The hydrodynamic expansion
in the presence of magnetic field has been studied in a longitudinal 
boost-invariant Bjorken picture by comparing the energy densities 
associated with the magnetic field and the fluid in event-by-event basis 
\cite{Umut:PRC89'2014,Roy:PRC92'2015}. Using the $(3 + 1)$-dimensional 
anomalous relativistic hydrodynamics \cite{Hirono:1412.0311}, it has been 
shown that the chiral magnetic effect is present 
in the charge dependent hadron azimuthal 
correlations. Similarly using the $(3+1)$-dimensional 
ideal hydrodynamics, the effect of spatially 
inhomogeneous magnetic field on the anisotropic 
expansion of the QGP in heavy ion collision has been studied 
in \cite{Pang:PRC93'2016}. There are also recent studies of the 
hydrodynamic expansion in the presence of magnetic field by parameterizing the 
magnetic nature of the matter, {\em i.e.} the EOS 
with nonzero magnetization \cite{Pu:PRD93'2016,Roy:PRC96'2017}. The 
abovementioned studies dealt with the conservation of energy-momentum tensor 
by incorporating the effect of magnetic field in it, known as 
relativistic magnetohydrodynamics. As our second aim, we will examine the 
same effects on the expansion through the EOS of the hot QCD matter, 
which is of paramagnetic nature, as an input 
in the hydrodynamical equation of motion, unlike the previous 
studies. Contrary to the above \cite{Pu:PRD93'2016,Roy:PRC96'2017}, 
we have derived the EOS from the free energy 
and checked its magnetic character, then plugged in to the hydrodynamics directly. 
As found earlier, the square of speed of sound ($c_s^2$) gets 
enhanced due to the effects of strong magnetic field, the energy density 
is consequently found to evolve faster than in the absence of strong magnetic 
field, implying that the matter now cools off faster.

The paper is organized in the following way. In section 2, we have 
evaluated the free energies due to quark and gluon contributions 
from the respective effective propagators and obtained the 
equation of state. Using the free energies thus obtained, we have
studied the magnetic properties of the 
QCD medium in the strong magnetic field by determining the 
magnetization and magnetic susceptibility in section 3. Then in 
section 4, we proceeded to observe the Bjorken expansion in the strong 
magnetic field regime. We concluded in section 5. Finally, in 
appendices A, B and C we have provided complete derivations of the quark 
self-energy, the quark contribution to the free energy and integrals in the 
presence of a strong magnetic field, respectively.

\section{Free energy}
The one-loop free energy (${\mathcal F}$) of a system of interacting quarks 
and gluons is the sum of free energies due to quarks and gluons, {\em i.e.} $\mathcal{F}=\mathcal{F}_q+\mathcal{F}_g$. In the coming subsections, we are 
going to discuss these contributions broadly.

\subsection{Quark contribution to the free energy}
The quark contribution (${\mathcal F_q}$) is calculated perturbatively 
in the fundamental representation of SU$(N_c)$ gauge theory 
by the functional determinant of the effective quark propagator, $S(P)$ ,
\begin{eqnarray}
\label{Q.F.E.} \mathcal{F}_q &=& 
N_cN_f\int\frac{d^4P}{(2\pi)^4}\ln\left[\det\left(S(P)
\right)\right]
,\end{eqnarray}
where $S(P)$ is obtained by the one-loop quark self-energy in a strong 
magnetic field through the Schwinger-Dyson equation,
\begin{equation}\label{E.P.}
S^{-1}(P)=S^{-1}_0(P)-\Sigma(P)
~.\end{equation}

The quark propagator was first obtained by Schwinger through the proper-time 
method \cite{Schwinger:PR82'1951} in the coordinate-space, 
\begin{equation}\label{S(X,Y)}
S(x,x^\prime)=\phi(x,x^\prime)\int\frac{d^4K}{(2\pi)^4}e^{-iK(x-x^\prime)}S(K)
~,\end{equation}
where the gauge-dependent phase factor, $\phi(x,x^\prime)$ is written 
in terms of path-representation of gauge fields,
\begin{equation}
\phi(x,x^\prime)=e^{i|q_f|\int^x_{x^\prime} A^\mu(\zeta)d\zeta_\mu}
~.\end{equation}
However, by making an appropriate gauge transformation, the phase factor 
can be gauged away. In fact, for a single fermion line, the phase factor in 
a symmetric gauge becomes unity, hence the propagator can also be written 
in the momentum-space as an integral over the Schwinger proper-time ($s$),
\begin{eqnarray}
\nonumber{S(K)} &=& i\int^{\infty}_0{ds}~\exp\left(isk_\parallel^2-{is}m_f^2
-\frac{ik_\perp^2\tan(|q_fB|s)}{|q_fB|}\right) \\ && \times\left\lbrace\left(m_f+
\gamma^\parallel\cdot{k}_\parallel\right)
\left(1+\gamma^1\gamma^2\tan(|q_fB|s)\right)
-\gamma^\perp\cdot{k}_\perp\left(1+\tan^2(|q_fB|s)\right)\right\rbrace
,\end{eqnarray}
where the notations are given by
\begin{eqnarray}
\nonumber&&k_\parallel\equiv(k_0,0,0,k_3),~~ k_\perp\equiv(0,k_1,k_2,0),
~~ \gamma^\parallel\equiv(\gamma^0,\gamma^3),~~ 
\gamma^\perp\equiv(\gamma^1,\gamma^2),~~ \\
&&g^{\mu\nu}_\parallel={\rm{diag}}(1,0,0,-1),~~ g^{\mu\nu}_\perp={\rm{diag}}(0,-1,-1,0), ~~ g^{\mu\nu}=g^{\mu\nu}_\parallel+g^{\mu\nu}_\perp \nonumber
.\end{eqnarray}
In terms of discretized Landau levels, $S(K)$ can be written 
in a discrete notation,
\begin{equation}\label{S(K)}
S(K)=ie^{-\frac{k^2_\perp}{|q_fB|}}\sum^\infty_{n=0}
(-1)^n\frac{D_n(|q_fB|,K)}{k^2_\parallel-m^2_f-2|q_fB|n}
~,\end{equation}
where $D_n$'s are expressed in terms of the Laguerre polynomials, $L_n$'s 
\cite{Tsai:PRD10'1974} as
\begin{eqnarray}\label{D_n(|q_fB|,K)}
\nonumber{D_n(|q_fB|,K)} &=& \left(\gamma^\parallel\cdot{k}_\parallel
+m_f\right)\left[\left(1-i\gamma^1\gamma^2\right)L_n\left(\frac{2k^2_\perp}{|q_fB|}\right)-\left(1+i\gamma^1\gamma^2\right)L_{n-1}\left(\frac{2k^2_\perp}
{|q_fB|}\right)\right] \\ && +4\gamma^\perp\cdot{k}_\perp{L^{(1)}_{n-1}}
\left(\frac{2k^2_\perp}{|q_fB|}\right)
~.\end{eqnarray}
In a strong magnetic field, only the lowest Landau level (LLL) ($n=0$) 
is occupied, hence the quark propagator (\ref{S(K)}) simplifies into
\begin{equation}\label{Q.P. in S.M.F.A.}
S_0 (K)=ie^{-\frac{k^2_\perp}{|q_fB|}}\frac{\left(\gamma^\parallel
\cdot{k}_\parallel+m_f\right)}{k^2_\parallel-m^2_f+i\epsilon}\left(1
-\gamma^0\gamma^3\gamma^5\right)
.\end{equation}

Since the gluons are electrically uncharged particles, they are not directly
affected by the Lorentz force, therefore the form of vacuum propagator 
for gluon remains the same, {\em i.e.} 
\begin{eqnarray}\label{G.P.}
D^{\mu\nu}_0 (Q)=\frac{ig^{\mu\nu}}{Q^2+i\epsilon}
~.\end{eqnarray}

In a thermal medium, both the Schwinger's proper-time propagator for quark 
(\ref{Q.P. in S.M.F.A.}) and the gluon propagator (\ref{G.P.}) get thermalized 
in the real-time formalism by the following $2\times2$ matrices,
\begin{eqnarray}
S(K)&=&\begin{pmatrix}n_F^{\prime\prime2} S_0(K)-n_F^{\prime2}S^*_0(K) & -
n_F^{\prime}n_F^{\prime\prime}(S_0(K)+S^*_0(K)) 
\\ n_F^{\prime}n_F^{\prime\prime}(S_0(K)+S^*_0(K)) & n_F^{\prime\prime2}S^*_0(K)-n_F^{\prime2}S_0(K)\end{pmatrix} \label{Q.P.M.}~, \\
D^{\mu\nu}(Q)&=&\begin{pmatrix}n_B^{\prime2}D^{*\mu\nu}_0(Q)+n_B^{\prime\prime2}D^{\mu\nu}_0(Q) & n_B^{\prime}n_B^{\prime\prime}(D^{\mu\nu}_0(Q)+D^{*\mu\nu}_0(Q)) 
\\ n_B^{\prime}n_B^{\prime\prime}(D^{\mu\nu}_0(Q)+D^{*\mu\nu}_0(Q)) & n_B^{\prime2}D^{\mu\nu}_0(Q)+n_B^{\prime\prime2}D^{*\mu\nu}_0(Q)\end{pmatrix} \label{G.P.M.}
~,\end{eqnarray}
respectively. The above notations are redefined 
in terms of quark ($n_F$) and gluon ($n_B$) equilibrium distribution functions 
as $n_F^\prime=\sqrt{n_F(k_0)}$, $n_F^{\prime\prime}=\sqrt{1-n_F(k_0)}$,
$n_B^{\prime}=\sqrt{n_B(q_0)}$ and $n_B^{\prime\prime}=\sqrt{1+n_B(q_0)}$.

For the description of a system in equilibrium quantum field theory, 
only the $11$-component of the self-energy 
matrix is sufficient, so we need
only the $11$-components of the quark and gluon propagators in a strongly 
magnetized thermal medium to calculate the respective effective propagators, 
which are therefore given by (from matrices (\ref{Q.P.M.}) and (\ref{G.P.M.})),
\begin{eqnarray}\label{11 Q.P.}
\nonumber{S_{11}}(K) &=& ie^{-\frac{k^2_\perp}{|q_fB|}}\left(\gamma^0k_0
-\gamma^3k_3+m_f\right)\left(1-\gamma^0\gamma^3\gamma^5\right) \\ 
&& \times\left[\frac{1}{k_{\parallel}^2-m_f^2+i\epsilon}+2\pi{i}n_F(k_0)
\delta\left(k_{\parallel}^2-m_f^2\right)\right], \\ \label{11 G.P.}
D^{\mu\nu}_{11}(Q) &=& ig^{\mu\nu}\left[\frac{1}{Q^2+i\epsilon}
-2\pi{i}n_B(q_0)\delta\left(Q^2\right)\right]
,\end{eqnarray}
respectively.
The temperature dependences are contained in the quark and gluon distribution 
functions,
\begin{eqnarray}
n_F(k_0) &=& \frac{1}{e^{\beta|k_0|}+1}~, \\ 
n_B(q_0) &=& \frac{1}{e^{\beta|q_0|}-1}
~,\end{eqnarray}
respectively with $T=1/\beta$.

Thus the $11$-component of one-loop 
quark self-energy (omitting the subscript ``$11$'') in terms of the 
$11$-components of the quark (\ref{11 Q.P.}) and gluon (\ref{11 G.P.}) 
propagators in a strong magnetic field is written as
\begin{eqnarray}\label{Quark self energy}
\nonumber\Sigma(P) &=& -\frac{4}{3} g^{2}
~i\int{\frac{d^4K}{(2\pi)^4}}\left[\gamma_\mu
{S_{11}(K)}\gamma^\mu{D_{11}(P-K)}\right] \\ 
&=& \nonumber\frac{4}{3}\frac{g^{2} ~ i}{(2\pi)^4}
\int{d^2k_{\perp}}{d^2k_{\parallel}e^{-\frac{k_{\perp}^2}{\mid{q_fB}\mid}}}\left[\gamma_\mu\left(\gamma^0k_0-\gamma^3k_3
+m_f\right)\left(1-\gamma^0\gamma^3\gamma^5\right)\gamma^\mu\right] 
\nonumber\\ && \nonumber\times\left[\frac{1}{k_{\parallel}^2-m_f^2
+i\epsilon}+2\pi{i}n_F(k_0)\delta\left(k_{\parallel}^2-m_f^2\right)\right] \\ 
&& \times\left[\frac{1}{(P-K)^2+i\epsilon}-2\pi{i}n_B(p_0-k_0)\delta
\left((P-K)^2\right)\right]
~,\end{eqnarray}
where the factor, $4/3$ represents the QCD colour factor and $g$ 
($g^2=4\pi\alpha_s (eB)$) is the running QCD coupling, that runs 
with the magnetic field because the magnetic field is the 
dominant energy scale of the thermal medium in a strong magnetic field 
($eB \gg T^2$), similar to the thermal medium in the absence of 
magnetic field where temperature is the dominant scale \cite{Ferrer:PRD91'2015,Andreichikov:PRL110'2013}. The trace of gamma 
matrices in eq. (\ref{Quark self energy}) is calculated as
\begin{eqnarray}
\gamma_\mu\left(\gamma^0k_0-\gamma^3k_3+m_f\right)\left(1
-\gamma^0\gamma^3\gamma^5\right)\gamma^\mu = -2\left[J_+\left(\gamma^0k_0
-\gamma^3k_3\right)-2m_f\right]
~,\end{eqnarray}
with $J_+=1+\gamma^0\gamma^3\gamma^5$. 

The quark self-energy (\ref{Quark self energy}) contains 
both the vacuum and medium contributions with single and product 
of two distribution functions. In the strong magnetic field (SMF) 
limit, due to the LLL approximation, the external quark momentum 
$P (p_\parallel, p_\perp)$ is assumed to be purely longitudinal 
\cite{Gusynin:PLB450'1999}, {\em i.e.} $p_\perp=0$. In addition, 
for the loop-momentum, both $k_\parallel^2$ and $k_\perp^2$ are 
much smaller than $|q_fB|$, so the form-factor, 
$e^{-k_{\perp}^2/|q_fB|}$ is approximated to be one. Thus the 
aforesaid artefacts of SMF limit facilitate to compute the 
real-part of the quark self-energy (in appendix A) as,
\begin{eqnarray}\label{T.Q.S.E.}
\nonumber\Sigma(p_{\parallel}) &=& \frac{{J_+}\left(\gamma^\parallel
\cdot{p}_\parallel\right)g^{2}}{12\pi^2}\left[-1
-\frac{|q_fB|}{m^2_f}\left\lbrace\ln\left(\frac{|q_fB|}{m^2_f}\right)
-1\right\rbrace\right] \\ && \nonumber+\frac{g^{2}m_f}{3\pi^2}\left[1
+\frac{{J_+}|q_fB|}{2m_f^2}\left\lbrace\ln\left(\frac{|q_fB|}{m^2_f}\right)
-1\right\rbrace\right] \\ && -\frac{2g^{2}m_f}{3\pi^2}
\ln\left(\frac{|q_fB|}{p_\parallel^2+m_f^2}\right)
\left[\ln\left(\frac{m_f}{\pi{T}}\right)+\gamma_E\right]
,\end{eqnarray}
which in turn evaluates the effective quark propagator 
from the Schwinger-Dyson equation (\ref{E.P.}).

Therefore the quark contribution to the free energy in a strong magnetic field 
is obtained from eq. (\ref{Q.F.E.}) (in appendix B),
\begin{eqnarray}\label{Free energy due to quarks}
\nonumber\mathcal{F}_q &=& -\frac{N_cN_f|q_fB|}{4}
\left[\frac{T^2}{3}+\int^{|q_fB|}_0\frac{dp^2_\parallel}{4\pi^2}\ln\left[\Bigg\lbrace\left(1-2{\eta}\right)^2
-\frac{1}{p^2_\parallel}\Bigg[m_f+2\Gamma+\xi\right.\right. \\ &+& \left.\left.\lambda\ln\left(\frac{|q_fB|}{p_\parallel^2+m_f^2}\right)\Bigg]^2\Bigg\rbrace\left\lbrace 1
-\frac{1}{p^2_\parallel}\left[m_f+\xi+\lambda\ln\left(\frac{|q_fB|}{p_\parallel^2+m_f^2}\right)\right]^2\right\rbrace\right]\right]
,\end{eqnarray}
where the new variables are defined as 
\begin{eqnarray}
&&\eta=-\frac{g^{2}}{12\pi^2}\left[1
+\frac{|q_fB|}{m^2_f}\left\lbrace\ln\left(\frac{|q_fB|}
{m^2_f}\right)-1\right\rbrace\right]\label{C},
~~ \Gamma=\frac{g^{2}}{6\pi^2}\frac{|q_fB|}{m_f}
\left\lbrace\ln\left(\frac{|q_fB|}{m^2_f}\right)-1\right\rbrace\label{D},\nonumber\\ 
&&\xi=\frac{g^{2}m_f}{3\pi^2}\label{E},
~~ \lambda=-\frac{2g^{2}m_f}{3\pi^2}
\left[\ln\left(\frac{m_f}{\pi{T}}\right)+\gamma_E\right]\label{W}\nonumber
.\end{eqnarray}

\subsection{Gluon contribution to the free energy}
To calculate the free energy due to the gluon 
contribution ($\mathcal{F}_g$) at finite temperature and finite 
magnetic field, we need to know the form of 
gluon self-energy tensor in the similar environment.
We will revisit here how to construct the general form of the gluon 
self-energy tensor ($\Pi^{\mu \nu}$) for an QCD medium at 
finite $T$ and finite $B$. Let us start with the vacuum, where the 
covariant form for the gluon self-energy tensor is given by
\begin{eqnarray}
\Pi^{\mu\nu}(p)=\Big(g^{\mu\nu}-\frac{p^\mu p^\nu}{p^2}\Big) \Pi (p^2)
~,\end{eqnarray}
where $\Pi^{\mu\nu}$ satisfies the transversality condition. Now bring the 
vacuum into the direct contact of a heat reservoir ($u^\mu$), so the 
Lorentz invariance is broken. Hence, a larger tensor basis is needed 
and the two orthogonal tensorial basis for finite temperature, 
$P^{\mu \nu}_T$ and $P^{\mu \nu}_L$ decompose the self-energy tensor \cite{Braaten:PRL64'1990,Braaten:PRD42'1990,Kobes:PRD45'1992} into
\begin{eqnarray}\label{g.s.e.(T)}
\Pi^{\mu\nu}(p_0,\mathbf{p})=\Pi_T P_T^{\mu\nu}(p_0,\mathbf{p}) + \Pi_L P_L^{\mu\nu}(p_0,\mathbf{p})
~,\end{eqnarray}
where
\begin{eqnarray}
P_T^{\mu\nu} &=& -g^{\mu\nu}+\frac{p_0}{\mathbf{p}^2}\left(p^\mu u^\nu+u^\mu p^\nu\right)-\frac{1}{\mathbf{p}^2}\left(p^\mu p^\nu+p^2u^\mu u^\nu\right),\label{transverse p.} \\ 
P_L^{\mu\nu} &=& -\frac{p_0}{\mathbf{p}^2}\left(p^\mu u^\nu+u^\mu p^\nu\right)+\frac{1}{\mathbf{p}^2}\left(\frac{p_0^2}{p^2} p^\mu p^\nu+p^2u^\mu u^\nu\right)
\label{longitudinal p.}
,\end{eqnarray}
are constructed in such a way that they satisfy the transversality condition 
- $p_\mu P^{\mu \nu}_{T,L}=0$. The subscripts $T$ and $L$ are
said to be transverse and longitudinal with respect to the three-momentum
($\mathbf{p}$), respectively. So the corresponding structure factors, 
$\Pi_T$ and $\Pi_L$ are known as transverse and longitudinal components of self-energy, which are obtained in hard thermal loop (HTL) approximation, where the hard scale 
for both quark and gluon loop momenta ($p$) is taken 
as the temperature ($p \sim T$), justifying the name HTL. As a result, 
both quark and gluon loops give the same result $\sim{\cal O } (g^2T^2)$ apart 
from flavor and colour factors, respectively.

Now we introduce the magnetic field in a direction, $b^\mu$, to the 
thermal medium. So, the remaining translation invariance is broken and 
much larger basis is needed, which can be constructed with the 
vectors, $p^\mu$, $u^\mu$, $b^\mu$ and the tensor, $g^{\mu \nu}$. Therefore, 
in addition to the earlier ones, two new orthogonal tensors, 
$P^{\mu \nu}_\parallel$ and $P^{\mu \nu}_\perp$ are constructed 
for finite magnetic field \cite{Hattori:PRD97'2018,Hattori:AP330'2013}, at 
least, for the leading-order perturbation theory,
\begin{eqnarray}
P_\parallel^{\mu\nu} &=& \frac{p_0p_3}{p_\parallel^2}\left(b^\mu u^\nu+u^\mu b^\nu\right)+\frac{1}{p_\parallel^2}\left(p_0^2 b^\mu b^\nu+p_3^2 u^\mu u^\nu\right),\label{parallel p.} \\  \nonumber P_\perp^{\mu\nu} &=& \frac{1}{p_\perp^2}\left[-p_\perp^2 g^{\mu\nu}+p_0\left(p^\mu u^\nu+u^\mu p^\nu\right)+p_3\left(p^\mu b^\nu+b^\mu p^\nu\right)-p_0p_3\left(b^\mu u^\nu+u^\mu b^\nu\right)\right. \\ && \left.\hspace{0.73 cm}-p^\mu p^\nu+\left(p_\perp^2-p_0^2\right)u^\mu u^\nu-\left(p_\perp^2+p_3^2\right)b^\mu b^\nu\right]
\label{perpendicular p.}
,\end{eqnarray}
by demanding the same transversality condition. The 
notations used in above equations are defined as follows:
\begin{eqnarray*}
&&u^{\mu}=(1,0,0,0),~b^\mu= (0,0,0,-1), ~ g^{\mu\nu}_\parallel= {\rm diag} (1,0,0-1), \\ &&g^{\mu\nu}_\perp={\rm diag} (0,-1,-1,0), ~ p^2_\parallel=p_0^2-p_3^2,~p^2_\perp=p_1^2+p_2^2
.\end{eqnarray*}
Therefore, the gluon self-energy tensor at finite $T$ and $B$ can be 
written as the superposition of finite $T$ 
(\eqref{transverse p.},\eqref{longitudinal p.}) and finite $B$ basis 
(\eqref{parallel p.},\eqref{perpendicular p.}),
\be\label{g.s.e.(T,eB)}
\Pi^{\mu\nu}=\Pi_T P_T^{\mu\nu} + \Pi_L P_L^{\mu\nu} + \Pi_\parallel P_\parallel^{\mu\nu} + \Pi_\perp P_\perp^{\mu\nu}
,\ee
with the two new structure factors, $\Pi_\parallel$ and $\Pi_\perp$ for 
finite $B$. 

Now the entire batch of structure factors, $\Pi_i$'s ($
{\rm i=T, L}, \parallel,\perp$) can be evaluated in HTL approximation
even in the presence of strong magnetic field
($eB \gg T^2$ as well as $eB \gg m^2$), but
with two hard scales, one is temperature and other is strong magnetic
field perceived by gluons and quarks, respectively. More specifically, the 
temperature remains the hard scale for the gluon-loop momentum because
gluons are not affected by the magnetic field but for the quark-loop momenta 
the hard scale is now replaced by the strong magnetic field ($\sqrt{eB}$)
only due to SMF limit ($eB \gg T^2$). Thus the gluon self-energy diagrams 
at finite $T$ and strong $B$ have been bifurcated into quark and gluon 
(ghost as well) loops, which give the thermal
and magnetic field contributions, respectively.

\subsubsection{Quark-loop contribution} 
The quark-loop contribution will be obtained through $\Pi_\parallel$ 
and $\Pi_\perp$. But for the strong magnetic field, the quarks are 
constrained to be in the lowest Landau levels and the dispersion relation
ensures the perpendicular component (with respect to the magnetic field)
to vanish, {\em i.e.} $p_\perp \approx 0$, which has been manifested by the 
vanishing structure factor, ($\Pi_\perp \approx 0$) \cite{Hattori:PRD97'2018,Ayala:1805.07344,Ayala:PRD98'2018}. Thus we are 
left to evaluate $\Pi_\parallel$.

In real-time formalism, using the LLL quark propagator 
(\ref{11 Q.P.}), the $11$-component of the gluon 
self-energy matrix (omitting the subscript ``$11$'') 
for the quark-loop is written as
\begin{eqnarray}\label{G.S.E. Expression}
\nonumber\Pi_q^{\mu\nu}(P) &=& -\frac{ig^{2}}{2}
\int\frac{d^4K}{(2\pi)^4}tr\left[\gamma^\mu{S_{11}(K)}
\gamma^\nu{S_{11}(K-P)}\right] \\ &=& \nonumber\frac{ig^{2}}
{2}\sum_f\int{\frac{d^2k_\perp}{(2\pi)^2}}
{\frac{d^2k_\parallel}{(2\pi)^2}}{tr}\left[\gamma^\mu
\left(\gamma^0k_0-\gamma^3k_3+m_f\right)\left(1-\gamma^0\gamma^3\gamma^5\right)\right. \\ && \left.\nonumber\times\gamma^\nu
\left(\gamma^0q_0-\gamma^3q_3+m_f\right)\left(1-\gamma^0\gamma^3\gamma^5\right)\right] e^{-\frac{k^2_\perp}{|q_fB|}}e^{-\frac{q^2_\perp}{|q_fB|}}
\\ && \nonumber\times\left[\frac{1}{k^2_\parallel-m^2_f+i\epsilon}+2\pi{i}n_F\left(k_0\right)\delta\left(k^2_\parallel-m^2_f\right)\right] 
\\ && \nonumber\times\left[\frac{1}{q^2_\parallel-m^2_f+i\epsilon}+2\pi{i}n_F\left(q_0\right)\delta\left(q^2_\parallel-m^2_f\right)\right] \\ 
&=& C_\perp(p_\perp)\Pi_q^{\mu\nu}(p_\parallel)
~,\end{eqnarray}
where the factor $1/2$ accounts for the trace over colour indices. 
Since the magnetic field decouples the transverse dynamics from the 
longitudinal one, the gluon self-energy in LLL is manifestly 
separated into transverse-momentum dependent and 
longitudinal-momentum dependent parts. The transverse-momentum 
dependent part is thus easily calculated as
\begin{eqnarray}\label{T.C.G.S.E.}
C_\perp(p_\perp) &=& \sum_f\frac{|q_fB|}{8\pi}
e^{-\frac{p^2_\perp}{2|q_fB|}}
.\end{eqnarray}
In the SMF limit, where $e^{-\frac{p^2_\perp}{2|q_fB|}} \approx 1$, the 
above equation is reduced into
\begin{eqnarray}\label{T.C.G.S.E.(eb)}
C_\perp &\simeq& \sum_f\frac{|q_fB|}{8\pi}
~.\end{eqnarray}
Next calculating the trace over gamma matrices, 
\begin{eqnarray}
L^{\mu\nu}=8\left[k^\mu_\parallel\cdot{q^\nu_\parallel}
+k^\nu_\parallel\cdot{q^\mu_\parallel}-g^{\mu\nu}_\parallel\left(k^\mu_\parallel\cdot{q}_{\parallel\mu}
-m^2_f\right)\right]
,\end{eqnarray}
the longitudinal-momentum dependent part can be written as the 
superposition of three terms,
\begin{eqnarray}\label{G.S.E.S.M.F.A.}
\Pi_q^{\mu\nu}(p_\parallel) &=& \Pi^{\mu \nu}_{q,V}(p_{\parallel})+\Pi^{\mu \nu}_{q,n}(p_{\parallel})+\Pi^{\mu \nu}_{q,n^2}(p_{\parallel})
~.\end{eqnarray}

The vacuum contribution in eq. (\ref{G.S.E.S.M.F.A.}) is decomposed as
\begin{eqnarray}\label{G.S.E.V.}
\Pi^{\mu \nu}_{q,V}(p_{\parallel}) &=& \frac{ig^{2}}{8\pi^2}
\int dk_0 dk_3\frac{L^{\mu\nu}}{(k_{\parallel}^2
-m_f^2+i\epsilon)(q_{\parallel}^2-m_f^2+i\epsilon)}
~,\end{eqnarray}
whose real part for the massless flavors yields to
\begin{equation}
\Re\Pi^{\mu\nu}_{q,V}(p_\parallel)=\frac{4g^{2}}{\pi}\left(g_{\parallel}^{\mu\nu}
-\frac{p_{\parallel}^{\mu}p_{\parallel}^{\nu}}{p_{\parallel}^2}
\right)
.\end{equation}
Thus after multiplying the transverse part 
\eqref{T.C.G.S.E.(eb)}, the real part of the longitudinal 
component (``00'' component, labelled as $\parallel$) of the 
vacuum part is simplified into
\begin{equation}\label{V.P.}
\Re\Pi^\parallel_{q,V} (p_0, p_3)=-\sum_f\frac{|q_fB|g^{2}p_{3}^2}{2\pi^2{p_{\parallel}^2}}
~.\end{equation}

Next the medium contribution with single 
distribution in eq. (\ref{G.S.E.S.M.F.A.}) is separated as
\begin{eqnarray}\label{G.S.E.S.D.}
\Pi^{\mu \nu}_{q,n}(p_{\parallel}) &=& -\frac{g^{2}}
{4\pi}\int dk_0 dk_3 L^{\mu\nu}\left[\frac{n_{F}(k_0)
\delta(k_{\parallel}^2-m_f^2)}{(q_{\parallel}^2
-m_f^2)}
+\frac{n_{F}(q_0)\delta(q_{\parallel}^2-m_f^2)}
{(k_{\parallel}^2-m_f^2)}\right]
,\end{eqnarray}
whose real part of ``$\parallel$'' component for massless 
flavors is calculated as
\begin{equation}\label{Massless}
\Re\Pi^\parallel_{q,n}(p_0,p_3)=-\frac{4g^{2}p_3^2}{{\pi}p_\parallel^2}
\left[\frac{T\ln(2)}{p_3}-1-\frac{T}{p_3}\ln\left(1
+e^{-\frac{p_3}{T}}\right)\right]
.\end{equation}
Thus multiplying the transverse part \eqref{T.C.G.S.E.(eb)}, the 
contribution to the real part containing single distribution function becomes
\begin{equation}\label{single d.f.}
\Re\Pi^\parallel_{q,n}(p_0,p_3)=-\sum_f\frac{|q_fB|g^{2}p_3^2}{2\pi^2p_\parallel^2}
\left[\frac{T\ln(2)}{p_3}-1-\frac{T}{p_3}\ln\left(1
+e^{-\frac{p_3}{T}}\right)\right]
.\end{equation}
The medium contribution involving product of two distribution functions in 
eq. (\ref{G.S.E.S.M.F.A.}),
\begin{eqnarray}\label{G.S.E.D.D.}
\Pi^{\mu \nu}_{q,n^2} (p_{\parallel}) &=& -\frac{ig^2}{2}
\int dk_0 dk_3 L^{\mu\nu}\left[n_{F}(k_0)n_{F}(q_0)
\delta(k_{\parallel}^2-m_f^2)
\delta(q_{\parallel}^2-m_f^2)\right]
,\end{eqnarray}
is purely imaginary, so it does not contribute to the real-part. Therefore we have
\begin{equation}\label{double d.f.}
\Re\Pi^\parallel_{q,n^2}(p_0,p_3) = 0
~.\end{equation}

So the vacuum (\ref{V.P.}) and medium 
(\eqref{single d.f.},\eqref{double d.f.}) 
contributions superpose together to give the 
real part of the gluon self-energy tensor 
due to quark-loop contribution,
\begin{eqnarray}\label{L.C.G.S.E.}
\nonumber\Re\Pi^\parallel_q(p_0,p_3) &=& \Re\Pi^\parallel_{q,V} (p_0, p_3)+\Re\Pi^\parallel_{q,n}(p_0,p_3)+\Re\Pi^\parallel_{q,n^2}(p_0,p_3) \\ &=& -\frac{g^2|q_fB|p_3}{2\pi^2p_\parallel^2}
\left[T\ln(2)-T\ln\left(1+e^{-\frac{p_3}{T}}\right)\right]
,\end{eqnarray}
which in the hard thermal loop approximation (external momentum, $p_3<T$), 
becomes temperature independent with the following form,
\begin{equation}\label{parallel1}
\Re \Pi^q_\parallel(p_0,p_3)=-\sum_f\frac{g^2|q_fB|p_3^2}{{4\pi^2}p_\parallel^2}
~.\end{equation}
Defining the square of the Debye mass due to the quark-loop
contribution only~\cite{Fukushima:PRD93'2016,MBB,Singh:PRD97'2018} as
\begin{eqnarray}\label{$m_D^2$}
m^2_{q,D} &=& \frac{g^{2}}{4\pi^2}\sum_{f}|q_{f}B|
~,\end{eqnarray}
the real part of the gluon self-energy due to the quark-loop becomes
\begin{equation}\label{parallel2}
\Re \Pi^q_\parallel(p_0,p_3)=-\frac{m_{q,D}^2p_3^2}{p_\parallel^2}
~.\end{equation}

Therefore the effective gluon propagator due to the quark-loop will be 
obtained from 
\begin{eqnarray}\label{P.G.P.}
\Delta_\parallel(P) &=& \frac{1}{\mathbf{p}^2+\Pi_\parallel}
~.\end{eqnarray}
Hence the quark-loop contribution to the gluon free energy 
is to be calculated from
\begin{eqnarray}\label{F.E.G.(eb)}
\mathcal{F}_g^{\rm quark~loop} &=& \frac{\left(N_c^2-1\right)}{2}\int\frac{d^4P}{(2\pi)^4}\ln\left(\mathbf{p}^2+\Pi_\parallel\right)
.\end{eqnarray}
In the hard region, momentum (squared) $\gg\Pi_\parallel$, so it 
is possible to expand $\mathbf{p}^2+\Pi_\parallel$ in powers of 
$\Pi_\parallel/\mathbf{p}^2$. After taking the leading orders 
(up to $\mathcal{O}\left(m_{q,D}^4\right)$) into account, 
eq. (\ref{F.E.G.(eb)}) can be written as
\begin{eqnarray}
\nonumber\mathcal{F}^{\rm quark~loop}_g &=& \frac{\left(N_c^2-1\right)}{2}
\int\frac{d^4P}{(2\pi)^4}\left[\ln\left(\mathbf{p}^2\right)
+\frac{\Pi_\parallel(p_\parallel)}{\mathbf{p}^2}-\frac{\Pi_\parallel^2(p_\parallel)}{2\mathbf{p}^4}\right] \\ 
&=& \frac{\left(N_c^2-1\right)}{2}\left[\int\frac{d^4P}{(2\pi)^4}\ln\left(\mathbf{p}^2\right)+\int\frac{d^4P}{(2\pi)^4}\frac{\Pi_\parallel(p_\parallel)}{\mathbf{p}^2}-\int\frac{d^4P}{(2\pi)^4}\frac{\Pi_\parallel^2(p_\parallel)}{2\mathbf{p}^4}\right]
.\end{eqnarray}
These integrals are all divergent. Using cutoff regularization method 
we have solved these integrals (appendix $C$) and found that the 
divergences nearly get canceled out through subtraction. Thus the 
gluon part of the free energy in this momentum region will not 
contribute to the total free energy. This fact also supports the 
claim in \cite{Hattori:PRD97'2018} that in the region where momentum (squared)  $\gg\Pi_\parallel$, all the self-energy corrections are negligible. 
However in the soft momentum region, $\Pi_\parallel=m_{q,D}^2$, so 
eq. (\ref{F.E.G.(eb)}) turns out to be
\begin{eqnarray}
\mathcal{F}_g^{\rm quark~loop} &=& \frac{\left(N_c^2-1\right)T}{2}\int\frac{d^3\mathbf{p}}{(2\pi)^3}\ln\left(\mathbf{p}^2+m^2_{q,D}\right)
.\end{eqnarray}
After solving we obtain the final value of the first part of the 
free energy as
\begin{eqnarray}\label{soft scale}
\mathcal{F}^{\rm quark~loop}_g &=& -\frac{\left(N_c^2-1\right)Tm^3_{q,D}}{12\pi}
~.\end{eqnarray}

\subsubsection{Gluon and ghost-loop contribution}
Now the structure factors, $\Pi_T$ and $\Pi_L$ are calculated 
using HTL approximation with temperature 
as the hard scale for gluon-loop momenta. Thus, in this 
case the effective gluon propagators for the transverse and 
longitudinal modes are obtained \cite{Andersen:PRD66'2002} as
\begin{eqnarray}
\label{T.G.P.} \Delta_T(P) &=& \frac{-1}{P^2+\Pi_T}~, \\ 
\label{L.G.P.} \Delta_L(P) &=& \frac{1}{\mathbf{p}^2+\Pi_L}
~,\end{eqnarray}
respectively. So the free energy due to the transverse and longitudinal 
components of the gluon propagator is calculated \cite{Andersen:PRD66'2002} as
\begin{eqnarray}\label{F.E.G.}
\mathcal{F}_g^{\rm gluon~loops} &=& \left(N_c^2-1\right)\left[
\int\frac{d^4P}{(2\pi)^4}\ln\left(P^2+\Pi_T\right)
+\frac{1}{2}\int\frac{d^4P}{(2\pi)^4}\ln\left(\mathbf{p}^2+\Pi_L\right)\right]
,\end{eqnarray}
where $\Pi_T$ and $\Pi_L$ are the components of gluon self-energy calculated 
from the gluon and ghost loops together using the hard thermal loop 
perturbation theory,
\begin{eqnarray}\label{T.G.S.E.}
\Pi_T &=& \frac{m^{\prime2}_{g,D}}{2}\frac{p^2_0}{\mathbf{p}^2}
+\frac{m^{\prime2}_{g,D}}{4}\frac{p_0}{\mathbf{p}}\left(1-\frac{p^2_0}{\mathbf{p}^2}\right)
\ln\left(\frac{p_0+\mathbf{p}}{p_0-\mathbf{p}}\right), \\
\Pi_L &=& m^{\prime2}_{g,D}-\frac{m^{\prime2}_{g,D}}{2}\frac{p_0}{\mathbf{p}}
\ln\left(\frac{p_0+\mathbf{p}}{p_0-\mathbf{p}}\right)
\label{L.G.S.E.}
,\end{eqnarray}
where the square of the Debye mass due to the 
gluon and ghost-loop is given by
\begin{eqnarray}\label{D.S.M.T.M.}
m^{\prime2}_{g,D} &=& \frac{g^{\prime2}T^2N_c}{3}
~,\end{eqnarray}
where the QCD coupling, $g^{\prime} (g^{\prime2}=4\pi\alpha_s^\prime(T))$, 
runs with the temperature. 

Thus, the free energy due to the gluon and ghost (\ref{F.E.G.}), up to 
$\mathcal{O}\left(m_{g,D}^{\prime4}\right)$ can be obtained 
\cite{Andersen:PRD66'2002} as
\begin{eqnarray}\label{part(1)}
\nonumber\mathcal{F}^{\rm gluon~loops}_g &=& -\left(N_c^2-1\right)\left[\frac{\pi^2T^4}{45}
-\frac{T^2m^{\prime2}_{g,D}}{24}+\frac{Tm^{\prime3}_{g,D}}{12\pi}\right. \\ && \left.\hspace{4.23 cm}+\frac{m^{\prime4}_{g,D}}{64\pi^2}\left\lbrace\ln\left(\frac{\Lambda}{4\pi{T}}\right)-\frac{7}{2}+\gamma_E+\frac{\pi^2}{3}\right\rbrace\right]
,\end{eqnarray}
where the renormalization scale ($\Lambda$) is set at $2\pi{T}$.

Therefore the quark \eqref{soft scale} and gluon-loop \eqref{part(1)} 
contributions together give the gluonic contribution to the free energy,
\begin{eqnarray}\label{T.G.S.S.F.E.P.}
\nonumber\mathcal{F}_g &=& -\left(N_c^2-1\right)\frac{Tm^3_{q,D}}{12\pi}-\left(N_c^2-1\right)
\left[\frac{\pi^2T^4}{45}
-\frac{T^2m^{\prime2}_{g,D}}{24}+\frac{Tm^{\prime3}_{g,D}}{12\pi}\right. \\ && \left.\hspace{4.3 cm}+\frac{m^{\prime4}_{g,D}}{64\pi^2}\left\lbrace\ln\left(\frac{\Lambda}{4\pi{T}}\right)-\frac{7}{2}+\gamma_E+\frac{\pi^2}{3}\right\rbrace\right]
,\end{eqnarray}
which along with the quark contribution \eqref{Free energy due to quarks} give the 
free energy of hot QCD matter in the presence of a strong magnetic field,
\begin{eqnarray}\label{F.E.}
\nonumber\mathcal{F} &=& -\frac{N_cN_f|q_fB|}{4}
\left[\frac{T^2}{3}+\int^{|q_fB|}_0\frac{dp^2_\parallel}{4\pi^2}\ln\left[\Bigg\lbrace\left(1-2\eta\right)^2
-\frac{1}{p^2_\parallel}\Bigg[m_f+2\Gamma+\xi\right.\right. \\ && \left.\left.\nonumber+\lambda\ln\left(\frac{|q_fB|}{p_\parallel^2+m_f^2}\right)\Bigg]^2\Bigg\rbrace\left\lbrace 1
-\frac{1}{p^2_\parallel}\left[m_f+\xi+\lambda\ln\left(\frac{|q_fB|}{p_\parallel^2+m_f^2}\right)\right]^2\right\rbrace\right]\right] \\ && \nonumber-\left(N_c^2-1\right)
\left[\frac{\pi^2T^4}{45}
-\frac{T^2m^{\prime2}_{g,D}}{24}+\frac{Tm^{\prime3}_{g,D}}{12\pi}+\frac{Tm^3_{q,D}}{12\pi}\right. \\ && \left.\hspace{3.9 cm}+\frac{m^{\prime4}_{g,D}}{64\pi^2}\left\lbrace\ln\left(\frac{\Lambda}{4\pi{T}}\right)-\frac{7}{2}+\gamma_E+\frac{\pi^2}{3}\right\rbrace\right]
.\end{eqnarray}

\subsection{Thermodynamic properties}
In the thermodynamic limit, the negative of the free energy gives the 
pressure ($P=-\mathcal{F}$) for a hot QCD medium, which is found to 
increase in the strong magnetic field regime \cite{Rath:JHEP1712'2017}, 
compared to the medium in the absence of magnetic field. However, the 
strong magnetic field restricts the dynamics of quarks to two dimensions, 
which causes a reduction of the phase space and results in the decrease 
of both entropy and energy densities. 
Finally we can obtain the equation of state of the hot QCD medium 
from the relation between the energy density and pressure,
\begin{eqnarray}
P = c_s^2 \varepsilon
~,\end{eqnarray}
by calculating the (square) speed of sound, $c_s^2$ from 
the free energy in the presence of a strong magnetic field (\ref{F.E.}),
\begin{eqnarray}\label{S3}
c_s^2=\frac{\partial P}{\partial \varepsilon}=\frac{
{\partial P}/{\partial T}}{{\partial \varepsilon}/{\partial T}}
~,\end{eqnarray}
which becomes a function of both temperature and magnetic field. To see the 
effects of the interplay between the temperature and (strong) magnetic 
field quantitatively, we have computed 
the (square) speed of sound ($c_s^2$) as a function of magnetic filed at 
different temperatures and {\em vice versa} in figures 1a and 1b, 
respectively and found an enhancement due to the presence of strong 
magnetic field. The above crucial observations on the thermodynamic 
observables as well as the equation of state may leave imprints on the 
magnetic properties of the medium and its subsequent 
hydrodynamic expansion in the ultrarelativistic heavy ion collisions.

\section{Magnetic properties in a strong magnetic field}
This section is devoted to explore the magnetic properties of the 
hot QCD matter in the presence of strong magnetic field. For this purpose 
we are going to calculate the lowest and the next-to-lowest order responses 
of the medium to the magnetic field, {\em i.e.} the magnetization and the magnetic 
susceptibility in the next subsections, respectively.

\begin{figure}[t]
\begin{center}
\begin{tabular}{c c}
\includegraphics[width=5.5cm]{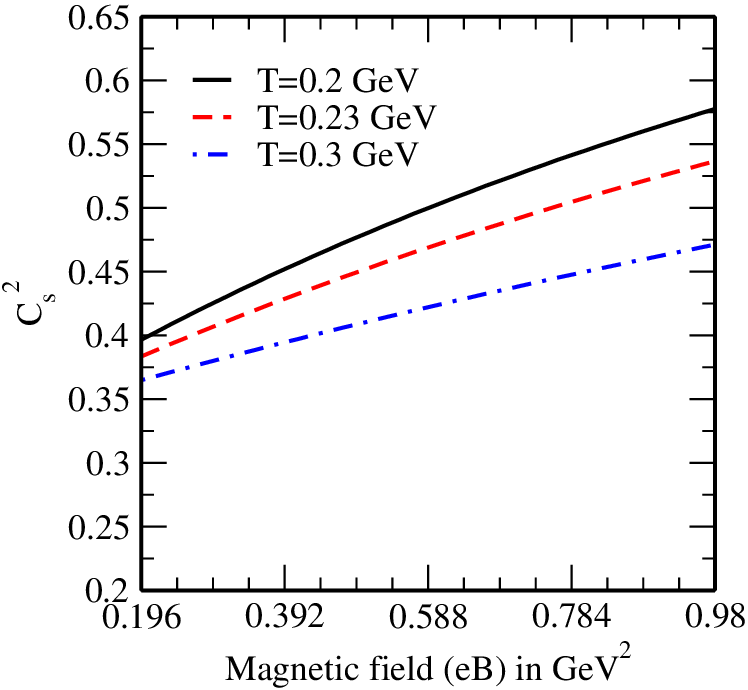}&
\hspace{1 cm}
\includegraphics[width=5.5cm]{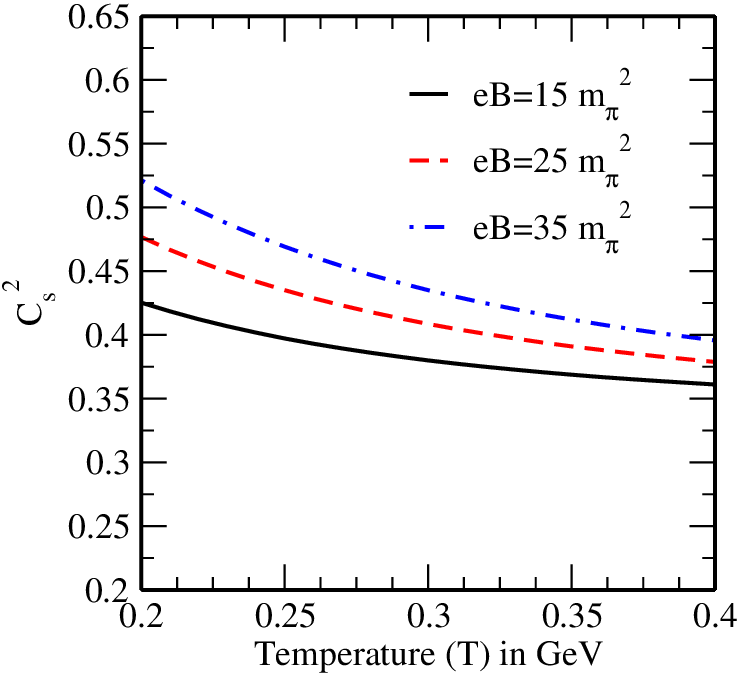}\\
a & b
\end{tabular}
\caption{The variation of square of the speed of sound 
with strong magnetic field at fixed temperatures (a) 
and the variation with temperature in the presence of strong 
magnetic fields (b).}
\end{center}

\end{figure}
\begin{figure}[t]
\begin{center}
\begin{tabular}{c c}
\includegraphics[width=5.5cm]{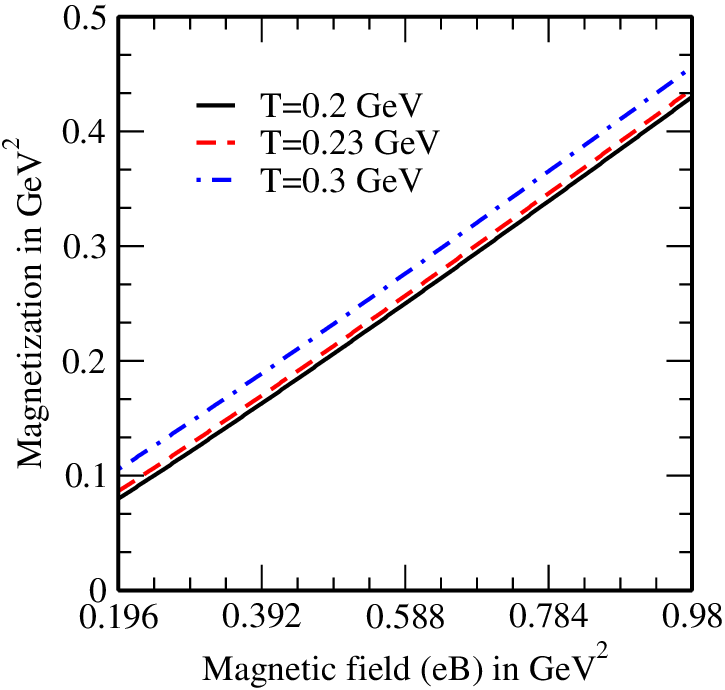}&
\hspace{1 cm}
\includegraphics[width=5.5cm]{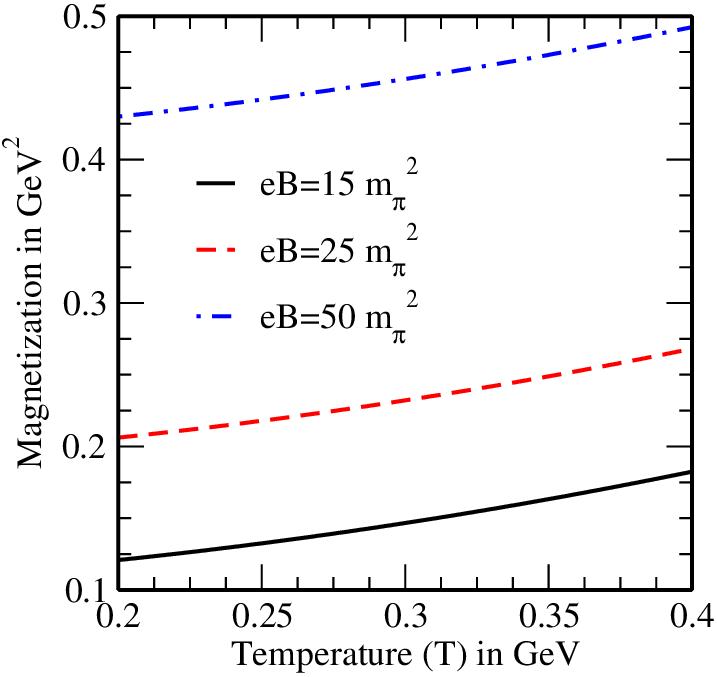}\\
a & b
\end{tabular}
\caption{The variation of magnetization 
as a function of the magnetic field at different temperatures 
(a) and as a function of the temperature at 
different magnetic fields (b).}
\end{center}
\end{figure}

\subsection{Magnetization}
Once the free energy (${\mathcal F}$) due to a medium of interacting quarks 
and gluons is known, the magnetization ($\mathcal{M}$) can be obtained from the 
following definition,
\begin{eqnarray}\label{Magnetization}
\mathcal{M} &=& -\frac{\partial\mathcal{F}}{\partial(eB)}
~.\end{eqnarray}

The final expression of $\mathcal{M}$ is given by
\begin{eqnarray}
\mathcal{M} &=& \nonumber\frac{N_cN_f}{4}
\left[\frac{T^2}{3}+\int^{|q_fB|}_0\frac{dp^2_\parallel}{4\pi^2}\ln\left[\Bigg\lbrace\left(1-2\eta\right)^2
-\frac{1}{p^2_\parallel}\Bigg[m_f+2\Gamma+\xi\right.\right. \\ && \left.\left.\nonumber+\lambda\ln\left(\frac{|q_fB|}{p_\parallel^2+m_f^2}\right)\Bigg]^2\Bigg\rbrace\left\lbrace 1
-\frac{1}{p^2_\parallel}\left[m_f+\xi+\lambda\ln\left(\frac{|q_fB|}{p_\parallel^2+m_f^2}\right)\right]^2\right\rbrace\right]\right] \\ 
&& \nonumber+\frac{N_cN_f|q_fB|}{16\pi^2}\int^{|q_fB|}_0dp_\parallel^2
~\left[\frac{1}{(1-2\eta)^2-\frac{1}{p_\parallel^2}\left\lbrace m_f+2\Gamma+\xi
+\lambda\ln\left(\frac{|q_fB|}{p_\parallel^2+m_f^2}\right)\right\rbrace^2}\right. \\ 
&& \left.\nonumber\times\left[\frac{g^{2}(1-2\eta)}{3\pi^2m^2_f}
\ln\left(\frac{|q_fB|}{m^2_f}\right)-\frac{2}{p_\parallel^2}
\left\lbrace m_f+2\Gamma+\xi+\lambda\ln\left(\frac{|q_fB|}{p_\parallel^2+m_f^2}\right)
\right\rbrace\right.\right. \\ && \left.\nonumber\left.\times\left\lbrace\frac{g^{2}}{3\pi^2m_f}\ln\left(\frac{|q_fB|}{m^2_f}\right)
+\frac{\lambda}{|q_fB|}\right\rbrace\right]-\frac{1}{1-\frac{1}{p_\parallel^2}\left\lbrace m_f+\xi+\lambda\ln\left(\frac{|q_fB|}{p_\parallel^2+m_f^2}\right)\right\rbrace^2}\right. \\ 
&& \left.\times\frac{2\lambda}{p^2_\parallel|q_fB|}\left\lbrace m_f+\xi+\lambda\ln\left(\frac{|q_fB|}{p_\parallel^2+m_f^2}\right)\right\rbrace\right]+\frac{\left(N_c^2-1\right)g^2Tm_{q,D}}{32\pi^3}
~.\end{eqnarray}

To analyse the magnetic behavior 
of QCD matter in a strong magnetic field, we have plotted 
the magnetization as a function of magnetic field at fixed temperatures 
$0.2$ GeV, $0.23$ GeV and $0.3$ GeV in figure 2a and {\em vice versa}
at magnetic fields, $15$ $m_\pi^2$, $25$ $m_\pi^2$ 
and $50$ $m_\pi^2$ in figure 2b. Similarly the magnetic susceptibility 
is also plotted as a function of temperature at different strong magnetic 
fields. In all these plots, we have set the ranges of magnetic field 
and temperature compatible to the strong magnetic field limit ($eB\gg{T^2}$).
We have observed that in the strong magnetic field limit, the magnetization 
increases almost linearly with the magnetic field and the 
increase is relatively larger for a hotter medium 
(in figure 2a). On the other hand the magnetization increases 
with the temperature too but the increase is meagre
(in figure 2b). The above (positive) magnetization can thus be explained 
by the decrease of 
free energy of the hot QCD matter in an ambient strong magnetic field, so it 
behaves as a paramagnetic medium. Minimization of the free energy 
induces a force, $F \equiv -\nabla{\mathcal{F}}$ which might affect 
the elliptic flow and causes a change in the shape of the QCD matter in the 
transverse plane of the noncentral heavy ion collision, known
as paramagnetic squeezing. Recent calculations in lattice 
\cite{Bonati:PRD89'2014,Levkova:PRL112'2014,Bali:PRL112'2014} and in 
hadron resonance gas model \cite{G:JHEP04'2013} also indicate that 
a hot QCD medium in a strong magnetic field is a paramagnetic one.

\begin{figure}[t]
\begin{center}
\begin{tabular}{c c}
\includegraphics[width=5.5cm]{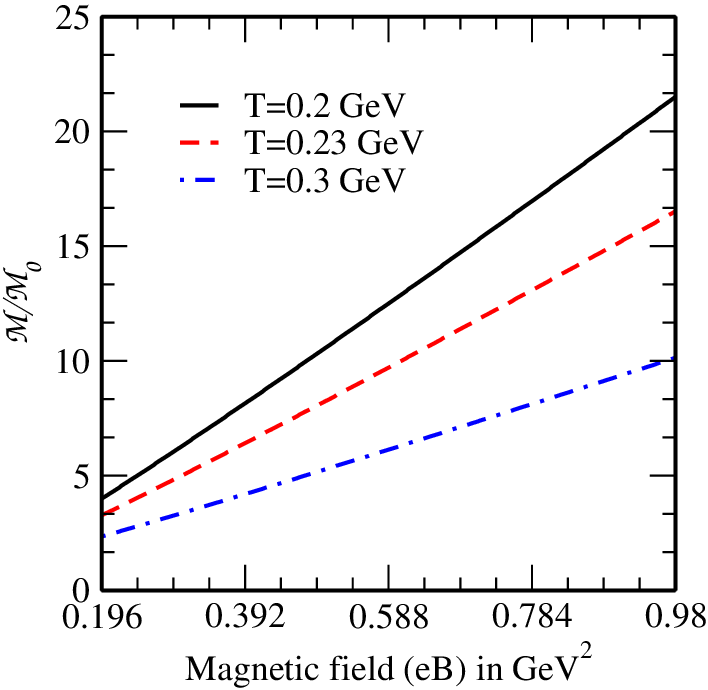}&
\hspace{1 cm}
\includegraphics[width=5.5cm]{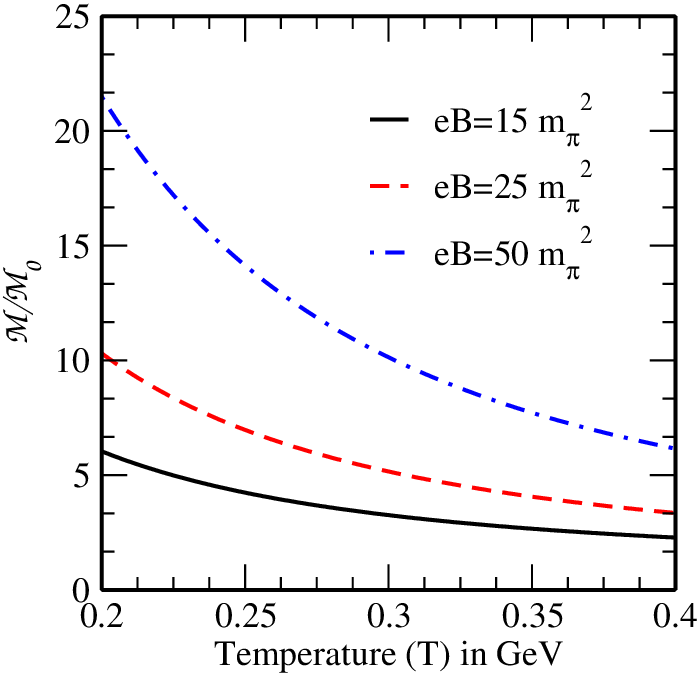}\\
a & b
\end{tabular}
\caption{Variation of magnetization normalized by its ideal 
value as a function of the magnetic field at different temperatures 
(a) and as a function of the temperature at different magnetic fields (b).}
\end{center}
\end{figure}

To see how the interactions among quarks and gluons affect the 
magnetization, we have computed the magnetization in units of its 
noninteracting value as a function of magnetic field at fixed temperatures 
in figure 3a and {\em vice versa} in figure 3b. We have observed that the 
magnetization gets enhanced due to the interaction and the enhancement is 
more pronounced with the increase of magnetic field. 
However the interaction has the opposite effect on the variation of 
magnetization with the temperature where the magnetization decreases 
with the temperature. This is due to the fact that as the temperature 
increases, the orientation of magnetic dipoles becomes more 
randomized, as a result the magnetization of the interacting quarks 
approaches beyond a temperature (greater than the magnetic interaction
energy) to its value attributed by the seemingly noninteracting quarks.

\subsection{Magnetic susceptibility}
The magnetic susceptibility ($\chi$) is a higher-order cumulant of free energy, 
which is obtained from the free energy as
\begin{eqnarray}\label{susceptibility}
\chi &=& -\frac{\partial^2\mathcal{F}}{\partial(eB)^2}
~.\end{eqnarray}

The final expression of $\chi$ is given by
\begin{eqnarray}
\chi &=& \nonumber\frac{N_cN_f}{8\pi^2}
\int^{|q_fB|}_0dp_\parallel^2~\left[\frac{1}{(1-2\eta)^2
-\frac{1}{p_\parallel^2}\left\lbrace m_f+2\Gamma+\xi
+\lambda\ln\left(\frac{|q_fB|}{p_\parallel^2+m_f^2}\right)\right\rbrace^2}\right. \\ 
&& \left.\nonumber\times\left[\frac{g^{2}(1-2\eta)}{3\pi^2m^2_f}
\ln\left(\frac{|q_fB|}{m^2_f}\right)
-\frac{2}{p_\parallel^2}\left\lbrace m_f+2\Gamma+\xi
+\lambda\ln\left(\frac{|q_fB|}{p_\parallel^2+m_f^2}\right)\right\rbrace\right.\right. \\ 
&& \left.\left.\nonumber\times\left\lbrace\frac{g^{2}}{3\pi^2m_f}
\ln\left(\frac{|q_fB|}{m^2_f}\right)
+\frac{\lambda}{|q_fB|}\right\rbrace\right]-\frac{1}{1-\frac{1}{p_\parallel^2}\left\lbrace m_f+\xi
+\lambda\ln\left(\frac{|q_fB|}{p_\parallel^2+m_f^2}\right)\right\rbrace^2}\right. \\ 
&& \left.\nonumber\times\frac{2\lambda}{p_\parallel^2|q_fB|}\left\lbrace m_f+\xi+\lambda\ln\left(\frac{|q_fB|}{p_\parallel^2+m_f^2}\right)\right\rbrace\right] \\ && 
\nonumber+\frac{N_cN_f|q_fB|}{16\pi^2}
\int^{|q_fB|}_0dp_\parallel^2\left[\frac{1}{(1-2\eta)^2
-\frac{1}{p_\parallel^2}\left\lbrace m_f+2\Gamma+\xi
+\lambda\ln\left(\frac{|q_fB|}{p_\parallel^2+m_f^2}\right)\right\rbrace^2}\right. \\ && 
\left.\nonumber\times\left[\frac{1}{2}\left\lbrace\frac{g^{2}}{3\pi^2m^2_f}\ln\left(\frac{|q_fB|}{m^2_f}\right)\right\rbrace^2+\frac{g^{2}(1-2\eta)}{3\pi^2m^2_f|q_fB|}-\frac{2}
{p_\parallel^2}\left\lbrace\frac{g^{2}}{3\pi^2m_f|q_fB|}-\frac{\lambda}{|q_fB|^2}\right\rbrace\right.\right. \\ 
&& \left.\left.\nonumber\times\left\lbrace m_f+2\Gamma+\xi
+\lambda\ln\left(\frac{|q_fB|}{p_\parallel^2+m_f^2}\right)\right\rbrace-\frac{2}
{p_\parallel^2}\left\lbrace\frac{g^{2}}{3\pi^2m_f}\ln\left(\frac{|q_fB|}{m^2_f}\right)+\frac{\lambda}{|q_fB|}\right\rbrace^2\right]\right. \\ 
&& \left.\nonumber-\frac{1}{\left[(1-2\eta)^2
-\frac{1}{p_\parallel^2}\left\lbrace m_f+2\Gamma+\xi
+\lambda\ln\left(\frac{|q_fB|}{p_\parallel^2+m_f^2}\right)\right\rbrace^2\right]^2}\right. \\ 
&& \left.\nonumber\times\left[-\frac{2}
{p_\parallel^2}\left\lbrace m_f+2\Gamma+\xi
+\lambda\ln\left(\frac{|q_fB|}{p_\parallel^2+m_f^2}\right)\right\rbrace\left\lbrace\frac{g^{2}}{3\pi^2m_f}\ln\left(\frac{|q_fB|}{m^2_f}\right)+\frac{\lambda}{|q_fB|}\right\rbrace\right.\right. \\ 
&& \left.\left.\nonumber+\frac{g^{2}(1-2\eta)}{3\pi^2m^2_f}\ln\left(\frac{|q_fB|}{m^2_f}\right)\right]^2+\frac{1}{\left[1-\frac{1}{p_\parallel^2}\left\lbrace m_f+\xi
+\lambda\ln\left(\frac{|q_fB|}{p_\parallel^2+m_f^2}\right)\right\rbrace^2\right]^2}\right. \\ 
&& \left.\nonumber\times\left[\frac{2\lambda}{p_\parallel^2|q_fB|^2}\left\lbrace m_f+\xi
+\lambda\ln\left(\frac{|q_fB|}{p_\parallel^2+m_f^2}\right)-\lambda\right\rbrace-\frac{2\lambda}{p_\parallel^4|q_fB|^2}\right.\right. \\ 
&& \left.\left.\nonumber\times\left\lbrace m_f+\xi
+\lambda\ln\left(\frac{|q_fB|}{p_\parallel^2+m_f^2}\right)+\lambda\right\rbrace\left\lbrace m_f+\xi
+\lambda\ln\left(\frac{|q_fB|}{p_\parallel^2+m_f^2}\right)\right\rbrace^2\right]\right] \\ 
&& +\frac{\left(N^2_c-1\right)g^4T}{256m_{q,D}\pi^5}
~.\end{eqnarray}

To investigate the response of magnetic field of the
hot QCD medium quantitatively, we have computed 
the magnetic susceptibility as a function of the temperature 
at different strong magnetic fields in figure 4, which comes
out to be positive and increases with both temperature and magnetic
field albeit the increase with the temperature is meagre.
Since the susceptibility remains positive therefore it reaffirms 
the paramagnetic nature of hot QCD medium in a strong magnetic field.

\begin{figure}[t]
\includegraphics[width=5.5cm]{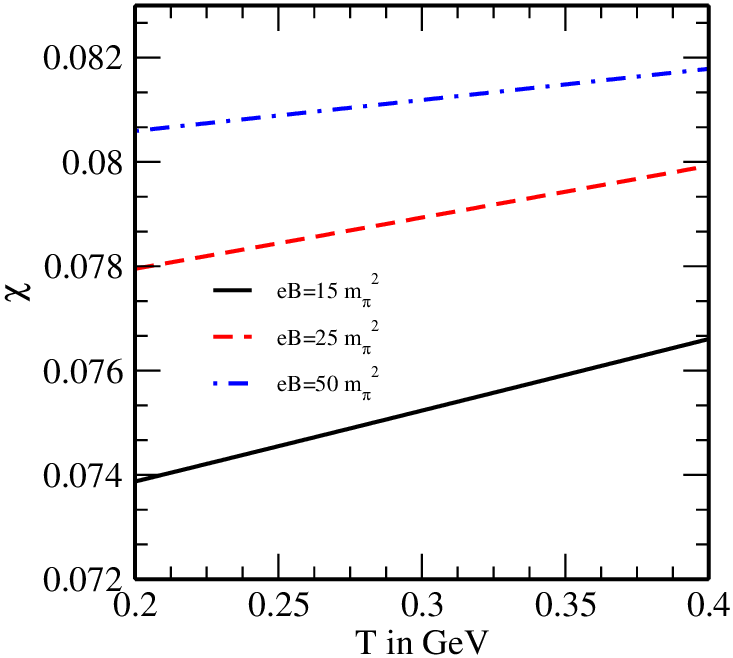}
\centering
\caption{The variation of magnetic susceptibility 
as a function of temperature at different magnetic fields.}
\end{figure}

Thus the free energy and its higher-order derivatives with respect 
to either magnetic field or temperature give the information 
about the magnetic and thermodynamic properties 
of a given medium in its different phases, respectively. 
In this work we have explored the 
magnetic behavior of QCD matter in a temperature range 
200-400 MeV in the presence of strong magnetic fields 
by calculating the magnetization and 
magnetic susceptibility, which are, respectively, given by the first- and 
second-order derivatives of the free energy with respect to the magnetic 
field. Similarly, in a pure thermal medium, the quark-number 
susceptibility can be obtained by differentiating the free energy with 
respect to the quark chemical potential to observe the baryon-number 
fluctuation, which facilitates to locate the critical point of phase transition 
\cite{Schaefer:PoS'2009,Koch:JPG35'2008}. Also the chiral susceptibility, 
which can be determined by the second-order derivative of the free energy with 
respect to the quark mass, helps to characterize the chiral phase transition 
in QCD \cite{He:PRD77'2008,Chang:PRC79'2009}.

\section{Bjorken expansion in the presence of a strong magnetic field}
In the vacuum the magnetic field decays very fast 
\cite{Kharzeev:NPA803'2008} so that it becomes tiny 
before a locally equilibrated thermal medium is created. Thus there 
may be some possible action of magnetic field on the static and dynamic 
properties of the medium produced in URHIC. However, some recent 
observations predict that the decay of magnetic field might be 
significantly slowed down due to the transport properties of the medium, 
{\em viz.} electrical conductivity~\cite{Umut:PRC89'2014,Tuchin:AHEP2013'2013,Conductivities}, 
as a result the magnetic field may be strong enough till the thermal 
medium is produced locally. This motivates to study the effect of strong 
magnetic field on the dynamics of the matter produced in the heavy ion collision.

The hydrodynamic expansion of the medium is given by the conservation of the 
energy-momentum tensor,
\begin{equation}
\partial_\mu T^{\mu \nu}=0
~,\end{equation}
where, in the absence of dissipative forces, it can be obtained as 
\begin{equation}
T^{\mu\nu}= (\varepsilon+P)u^\mu u^\nu - g^{\mu \nu} P ~,
\label{tmu}
\end{equation}
where $\varepsilon$ and $P$ are the energy density and the pressure of the 
medium, respectively. These are obtained from the free energy 
of the thermal QCD medium in a strong magnetic field, which is shown 
to behave as a paramagnetic medium, {\em i.e.} having positive magnetization.

Therefore the ideal fluid gives the longitudinal boost-invariant Bjorken expansion,
\begin{equation}
\frac{d \varepsilon}{d \tau} =-\frac{\varepsilon+P}{\tau}
~.\label{bj}
\end{equation}
Using the equation of state ($P=c_s^2 \varepsilon$) 
for thermal QCD medium in a strong magnetic field, the above first-order 
differential equation can be solved to give the evolution of 
energy density as a function of the proper-time,
\begin{eqnarray}
\label{eq0}
\varepsilon(\tau) \tau^{1+c_s^2}= \varepsilon(\tau_i)\tau_i^{1+c_s^2},
\end{eqnarray}
where $c_s^2$ is taken from the eq.~(\ref{S3}) as a function of
both temperature and magnetic field, whereas in usual hydrodynamics,
$c_s^2$ is either taken as a constant ($1/3$) for noninteracting equation
of state or taken as a function of temperature only (for baryonless
matter). {\em It is very important to mention here} that the effect 
of magnetic field on the expansion dynamics enters through the thermodynamic 
equation of state, unlike the usual magnetohydrodynamics for the static as 
well as dynamic magnetic fields, where the energy-momentum tensor 
($T^{\mu\nu}$) is decomposed into the matter ($T_M^{\mu\nu}$) and the 
(external) field ($T_B^{\mu\nu}$) contributions, as in 
\cite{Pu:PRD93'2016,Roy:PRC96'2017}. The latter comes as a product of the 
canonically conjugate variables. Here these are magnetization and magnetic 
field for which they have taken some specific values for them. On the 
contrary we have started with the EoS for the hot QCD medium in the presence 
of a strong magnetic field, where both pressure and energy density are 
functions of temperature and magnetic field and thus it automatically 
guarantees the effect of (strong) magnetic field into the EoS, confirming 
the paramagnetic nature. As a result, in our case, the energy-momentum tensor 
(that contains the matter part) itself incorporates the 
effects of temperature, magnetic field and initial magnetization, so, there 
is no need to include the magnetic field part ($T_B^{\mu\nu}$) in the 
energy-momentum tensor further.

The proper-time evolution of energy density 
can be translated, by the equation of state ($\varepsilon \sim T^4$), 
into the evolution of temperature with (proper) time, 
known as cooling law. As an example, using the noninteracting 
equation of state for a thermal QCD medium, eq. (\ref{eq0}) 
can be converted into
\begin{eqnarray}
\label{eq1}
T^3 \tau =T_i^3 \tau_i
~.\end{eqnarray}

The corrections to the Bjorken expansion for ideal fluid
can be incorporated by adding the shear stress tensor, 
$\pi^{\mu\nu}$ to the ideal case (\ref{tmu}), 
\begin{equation}
T^{\mu\nu}= (\varepsilon+P)u^\mu u^\nu - g^{\mu \nu} P +\pi^{\mu\nu} ~,
\label{tmun1}
\end{equation}
which can be expressed in terms of the velocity-gradient,
\be
\pi^{\mu\nu}=\eta \left( \nabla^\mu u^\nu + \nabla^\nu u^\mu -
\frac{2}{3} \nabla^{\mu \nu} \nabla^\rho u_\rho \right)
~,\ee
where $\eta$ is the shear viscosity and 
$\nabla^\mu=\nabla^{\mu \nu} \partial_\nu$ with $\nabla^{\mu \nu}=g^{\mu \nu} - u^\mu u^\nu$.

In a first-order viscous hydrodynamics, the shear stress tensor is written in 
terms of the first-order symmetrized velocity gradient ($ \langle \nabla^\mu 
u^\nu \rangle$),
\be
\pi^{\mu\nu}=\eta \langle 
\nabla^\mu u^\nu \rangle
~,\ee
whereas for a second-order viscous hydrodynamics, Israel-Stewart theory 
\cite{Israel:1979wp} modifies the equation of motion for the ideal fluid 
(\ref{bj}) into \cite{Muronga:PRL88'2002,Baier:2006um},
\be
\frac{d \varepsilon} {d \tau}=-\frac{1}{\tau}(\varepsilon+P-\Phi)
\label{eq5}
~,\ee
where the shear stress ($\Phi$) will asymptotically be 
reduced into its first-order value after the relaxation time ($\tau_\pi$)
is elapsed. In the Navier-Stokes limit,  the 
one-dimensional boost-invariant expansion gives~\cite{Muronga:PRL88'2002}, 
\begin{equation}
\Phi = \frac{4\eta}{3\tau}
\label{eq:eta-zeta}
~.\end{equation}

Thus substituting the value of $\Phi$ in eq. (\ref{eq5}), the 
Bjorken longitudinal expansion for dissipative fluid 
can be obtained as
\be
\frac{d \varepsilon} {d \tau}+\frac{\varepsilon+P}{\tau}=
\frac{4\eta}{3\tau^2}
~.\label{eqbj2}
\ee
The solution of eq. (\ref{eqbj2}) with the EoS $P=c_s^2 \varepsilon$, 
is determined as
\begin{eqnarray}
\label{eqs1}
\varepsilon(\tau) \tau^{1+c_s^2}+\left\lbrace\frac{4a_f T_{i}^3 \tau_{i}}{3}\right\rbrace\left\lbrace\frac{\eta}{s}\right\rbrace \frac{\tau^{1+c_s^2}}{{\tilde{\tau}}^2}
&=&\varepsilon(\tau_i)\tau_i^{1+c_s^2}
+\left\lbrace\frac{4a_f T_{i}^3 \tau_{i}}{3}\right\rbrace\left\lbrace\frac{\eta}{s}\right\rbrace \frac{\tau_i^{1+c_s^2}}{ {\tilde{\tau_i}}^2} 
~,\end{eqnarray}
where $a_f=(16+21N_f/2)\pi^2 /90$, ${\tilde{\tau}}^2$ 
and ${\tilde{\tau}}_i^2$ denote $(1-c_s^2)\tau^2$ 
and $(1-c_s^2)\tau_i^2$, respectively. In the above solution, the first 
and the second terms in both sides represent the contributions due to the 
zeroth-order expansion and due to the viscous corrections, respectively. 
In the present work we have used the shear viscosity to entropy ratio, 
$\eta/s$ as $0.08$ from the AdS/CFT \cite{ADS-kovtun-dtson-PRL942005} 
calculation and $0.3$ from the perturbative QCD \cite{pqcd} calculation. 
In our calculations we have also set $\tau_i=0.25$ fm/c, 
$T_i=0.39$ GeV for centrality $(40 - 50)$ \% at energy $\sqrt{s_{NN}}=200$ 
GeV from the RHIC Au$+$Au data \cite{Kisiel:PRC79'2009}, and through the 
equation of state the initial temperature gives 
$\varepsilon_i=36.833$ GeV/${\rm{fm}}^3$.

Similar to the cooling law (\ref{eq1}) in ideal hydrodynamics, the 
cooling law in a first-order dissipative hydrodynamics can also be 
obtained in terms of the $\eta/s$ ratio,
\begin{equation}
T(\tau)=T_i\left(\frac{\tau_i}{\tau}\right)^{1/3}\left[1+\frac{2\eta}{3\tau_i{T_i}s}\left\lbrace1-\left(\frac{\tau_i}{\tau}\right)^{2/3}\right\rbrace\right]
.\end{equation}

\begin{figure}[t]
\begin{center}
\begin{tabular}{c c c}
\includegraphics[width=5cm]{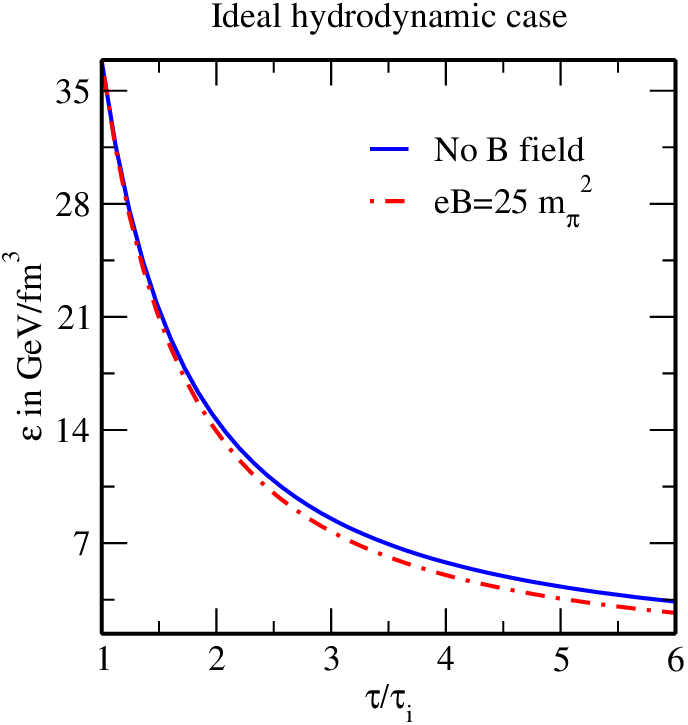}&
\hspace{0.09 cm}
\includegraphics[width=5cm]{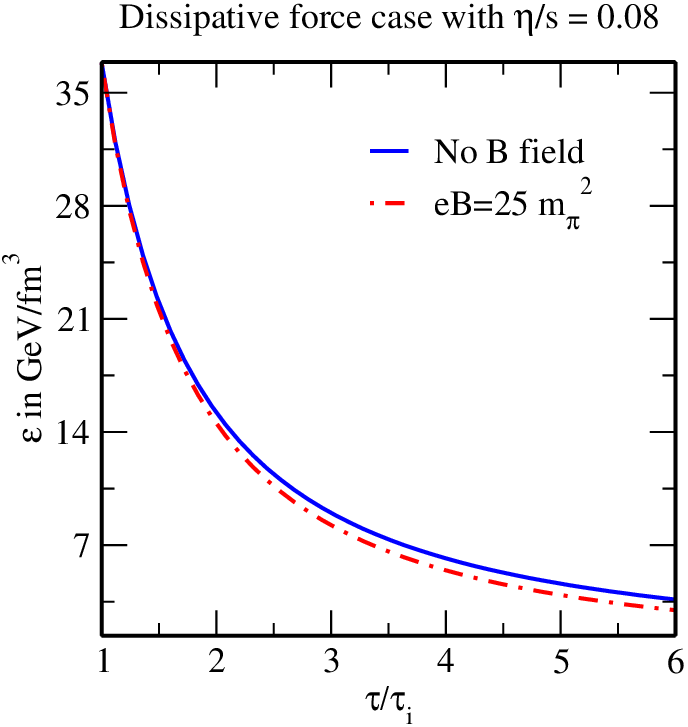}&
\hspace{0.09 cm}
\includegraphics[width=5cm]{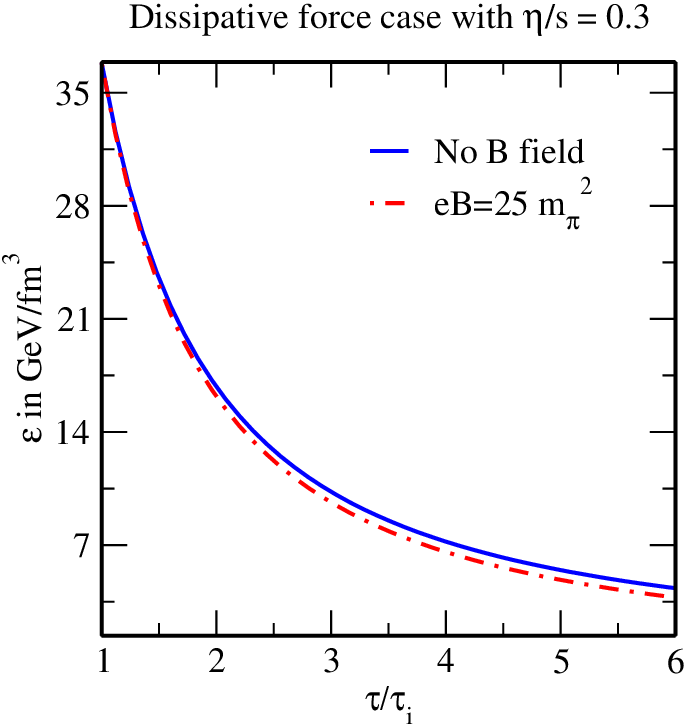}\\
a & b & c
\end{tabular}
\caption{Evolution of the energy density in case of (a) ideal 
hydrodynamics and in the presence of dissipative forces 
with the ratio, (b) $\frac{\eta}{s}=0.08$ and (c) $\frac{\eta}{s}=0.3$,
respectively.}
\end{center}
\end{figure}

To know how an ambient strong magnetic field could affect the 
hydrodynamic expansion quantitatively, we have observed the evolution 
of the energy density $(\varepsilon)$ of a hot QCD medium 
with the proper-time ($\tau$) in the absence (solid lines) and in the 
presence of strong magnetic field (dotted-dashed lines) in figures 5a 
and 5b and 5c for the ideal and dissipative fluids, respectively. The 
energy density always decreases with the proper time, but in the 
presence of strong magnetic field, the decrease of $\varepsilon$ with $\tau$ 
becomes faster as compared to the case in the absence of magnetic field. This 
observation can be envisaged from our earlier observation on the equation of 
state (figure 1), where the speed of sound gets enhanced due to the presence 
of strong magnetic field. Similar behavior in the energy density as a 
function of proper time is also noticed in the presence of dissipative 
forces, {\em i.e.} with the nonzero shear viscosity to entropy ratio ($\eta/s$).

\section{Conclusions}
We have studied the response of strong magnetic field on a hot 
QCD medium by estimating the magnetization and the 
magnetic susceptibility from its one-loop 
free energy evaluated in the strong magnetic field
limit ($eB \gg T^2$), which may be realized in the non-central events 
of ultra-relativistic heavy 
ion collisions. Both the magnetization and magnetic susceptibility 
have been emanated from the quark and gluon contributions through 
their respective one-loop self-energies, where the 
quark contribution is largely affected by the 
magnetic field unlike the gluon contribution. The 
positive values of both magnetization and magnetic 
susceptibility, which in turn increase with the magnetic field, 
have affirmed the paramagnetic nature of the hot 
QCD medium in the presence of a strong magnetic field. In addition, 
we have explored the effect of strong magnetic field on the evolution 
of the QCD medium both in the presence and absence of dissipative 
forces, where the evolution of the energy density with proper time is 
observed to get faster than that in a pure thermal medium in the absence 
of magnetic field.

\section{Acknowledgment}
We would like to thank Prof. Somendra Mohan Bhattacharjee for fruitful 
discussions during the preparation of this manuscript.

\appendix
\appendixpage
\addappheadtotoc
\begin{appendices}
\renewcommand{\theequation}{A.\arabic{equation}}
\section{Quark self-energy in a strong magnetic field}
In the presence of a magnetic field, the motion of a 
quark perpendicular to the direction of magnetic 
field is quantized in terms of Landau levels. In the SMF limit, 
{\em i.e.} in the LLL approximation 
$k_\perp^2\ll|q_fB|$ and this leads 
$\exp(-k_{\perp}^2/|q_fB|)\approx 1$. Also the upper 
limits of the momentum integrals are set at $|q_fB|$ 
and $p_\perp \approx 0 $ \cite{Gusynin:PLB450'1999}, so we have
\begin{eqnarray}(P-K)^2 \simeq \left(p_\parallel-k_\parallel\right)^2
-k_\perp^2
~.\end{eqnarray}
Therefore the quark self-energy (\ref{Quark self energy}) 
becomes a function of the longitudinal component of 
momentum only with the following form,
\begin{eqnarray}\label{Total sigma}
\nonumber\Sigma(p_{\parallel}) &=& \frac{-8g^{2}i}{3(2\pi)^4}\int{d^2k_{\perp}{d^2k_\parallel}}\left[J_+\left(\gamma^0k_0
-\gamma^3k_3\right)-2m_f\right] 
\\ && \nonumber\times\left[\frac{1}{k_{\parallel}^2-m_f^2+i\epsilon}+2\pi{i}n_F(k_0)\delta(k_{\parallel}^2-m_f^2)\right] 
\\ && \nonumber\times\left[\frac{1}{(p_\parallel-k_\parallel)^2
-k_\perp^2+i\epsilon}-2\pi{i}n_B(p_0-k_0)
\delta((p_\parallel-k_\parallel)^2-k_\perp^2)\right] \\ 
&\equiv& \Sigma_V(p_{\parallel})+\Sigma_n(p_{\parallel})
+\Sigma_{n^2}(p_{\parallel})
~,\end{eqnarray}
where $\Sigma_V(p_{\parallel})$, $\Sigma_n(p_{\parallel})$ and 
$\Sigma_{n^2} (p_{\parallel})$ denote the vacuum part, 
medium parts containing single distribution function and 
product of two distribution functions, respectively.

\subsection{Vacuum part}

The vacuum part of the quark self-energy 
from eq. (\ref{Total sigma}) can be written as
\begin{eqnarray}
\nonumber\Sigma_V(p_{\parallel}) &=& \frac{-8g^{2}i}{3(2\pi)^4}
\int{d^2k_{\perp}{d^2k_\parallel}}\left[J_+\left(\gamma^0k_0-\gamma^3k_3\right)
-2m_f\right] \\ && \times\left[\frac{1}{k_{\parallel}^2-m_f^2
+i\epsilon}\right]\left[\frac{1}{(p_\parallel-k_\parallel)^2
-k_\perp^2+i\epsilon}\right]
.\end{eqnarray}
Using the identity,
\begin{equation}\label{identity}
\frac{1}{x\pm{y}\pm{i\epsilon}}={\rm{P}}\left(\frac{1}
{x\pm{y}}\right)\mp{i\pi{\delta(x\pm{y})}}
~,\end{equation}
we obtain the real part with the following form,
\begin{eqnarray}\label{$Sigma_v$}
\nonumber\Sigma_V(p_{\parallel}) &=& \frac{-8g^{2}}{3(2\pi)^4}
\int{d^2k_{\perp}}\int{d^2k_{\parallel}}\left[J_+\gamma^\parallel\cdot{k_\parallel}
-2m_f\right]\frac{i}{\left[k_{\parallel}^2
-m_f^2\right]\left[(p_\parallel-k_\parallel)^2-k_\perp^2\right]} \\ 
&\equiv & \frac{-8g^{2}}{3(2\pi)^4}\int{d^2k_{\perp}}I
~.\end{eqnarray}
Using Feynman parametrization method and Wick rotation, integration $I$ can be 
solved with the change of variable from $k_\parallel$ to $k^\prime$ through the 
replacement $k_\parallel-zp_\parallel=k^\prime$,
\begin{eqnarray}\label{K.P. integration}
I &=& -\int^1_0~dz~d^2k^\prime~\frac{zJ_+\gamma^\parallel\cdot{p_\parallel}-2m_f}{\left[{k^\prime}^2-z(1-z)p^2_\parallel+zk^2_\perp+(1-z)m^2_f\right]^2} \nonumber\\
&=& \frac{\pi}{|q_fB|}\int^1_0dz\left(zJ_+\gamma^\parallel
\cdot{p}_\parallel-2m_f\right)+\pi\int^1_0dz\frac{-zJ_+\gamma^\parallel
\cdot{p}_\parallel+2m_f}{-z(1-z)p^2_\parallel+zk^2_\perp+(1-z)m^2_f} 
\nonumber\\ 
&\equiv& I^a+I^b
~,\end{eqnarray}
After solving the integral $I^a$ analytically, we get
\begin{eqnarray}
{I^a} &=& \frac{\pi}{2|q_fB|}\left(J_+\gamma^\parallel\cdot{p}_\parallel-4m_f\right)
.\end{eqnarray}
The integral $I^b$ is solved by expanding it in a Taylor series around the mass-shell 
condition - $\gamma^\parallel\cdot{p}_\parallel=m_f$ in a strong magnetic 
field,
\begin{eqnarray}
I^b=A+B\left(\gamma^\parallel\cdot{p}_\parallel-m_f\right)
+C\left(\gamma^\parallel\cdot{p}_\parallel-m_f\right)^2+\cdots
~.\end{eqnarray}
Up to first order (dropping the higher-order terms), $I^b$ is given by 
\begin{eqnarray}\label{I_2}
I^b=A+B\left(\gamma^\parallel\cdot{p}_\parallel-m_f\right),
\end{eqnarray}
where the coefficients, $A$ and $B$ are calculated as follows,
\begin{eqnarray}\label{A integration}
A &=& \left.I^b\right\vert_{\gamma^\parallel
\cdot{p}_\parallel=m_f} \nonumber\\ 
&=& -\pi{m_f}\int^1_0dz\frac{zJ_+-2}{m^2_f(1-z)^2+zk^2_\perp}~, \\ 
\label{B integration} B &=& \left.\frac{\partial{I^b}}{\partial\left(\gamma^\parallel 
\cdot{p}_\parallel\right)}\right
\vert_{\gamma^\parallel\cdot{p}_\parallel=m_f} \nonumber\\
&=& -{\pi}\int^1_0dz\frac{zJ_+}{(1-z)^2m^2_f+zk^2_\perp}-\pi\int^1_0dz
\frac{2m^2_fz(1-z)(zJ_+-2)} {\left\lbrace(1-z)^2 m^2_f+zk^2_\perp\right\rbrace^2}
~.\end{eqnarray}
The Feynman integrations in equations (\ref{A integration}) and 
(\ref{B integration}) yield the expressions for the 
coefficients $A$ and $B$ which for masses of two light flavors 
$u$ and $d$, are simplified into,
\begin{eqnarray}\label{A}
A &\simeq& -\frac{\pi{J_+}}{2m_f}\ln\left(\frac{k^2_\perp}{m^2_f}\right), \\ 
\label{B} B & \simeq& \frac{\pi{J_+}}{2m^2_f}\ln\left(\frac{k^2_\perp}{m^2_f}\right)
.\end{eqnarray}
Therefore the integral, $I^b$ takes the following form,
\begin{eqnarray}
I^b = -\frac{\pi{J_+}}{m_f}\ln\left(\frac{k^2_\perp}{m^2_f}\right)+\frac{\pi{J_+}}{2m^2_f}\left(\gamma^\parallel\cdot{p}_\parallel\right)\ln\left(\frac{k^2_\perp}{m^2_f}\right)
.\end{eqnarray}
Now the $k_\parallel$-integration ($I$) in 
eq. (\ref{K.P. integration}) turns out to be
\begin{eqnarray}
I = \frac{\pi}{2|q_fB|}\left({J_+}\gamma^\parallel\cdot{p}_\parallel
-4m_f\right)-\frac{\pi{J_+}}{m_f}\ln\left(\frac{k^2_\perp}{m^2_f}\right)+\frac{\pi{J_+}}{2m^2_f}\left(\gamma^\parallel\cdot{p}_\parallel\right)\ln\left(\frac{k^2_\perp}{m^2_f}\right)
.\end{eqnarray}
Thus the real part of the vacuum contribution in eq. (\ref{$Sigma_v$}) becomes
\begin{eqnarray}\label{QSE in vacuum}
\nonumber\Sigma_V(p_{\parallel}) &=& \frac{-8g^{2}\pi}{3(2\pi)^4} \int_0^{|q_fB|} dk_{\perp}^2I \\ &=& \nonumber\frac{{J_+}\left(\gamma^\parallel
\cdot{p}_\parallel\right)g^{2}}{12\pi^2}\left[-1
-\frac{|q_fB|}{m^2_f}\left\lbrace\ln\left(\frac{|q_fB|}{m^2_f}\right)
-1\right\rbrace\right] \\ && +\frac{g^{2}m_f}{3\pi^2}\left[1
+\frac{{J_+}|q_fB|}{2m_f^2}\left\lbrace\ln\left(\frac{|q_fB|}{m^2_f}\right)
-1\right\rbrace\right]
.\end{eqnarray}

\subsection{Medium part}

The medium part, $\Sigma_{n}(p_{\parallel})$, which depends 
on the quark and gluon distribution functions linearly is given by
\begin{eqnarray}\label{S.D.F.P}
\nonumber\Sigma_n(p_{\parallel}) &=& \frac{8g^{2}}{3(2\pi)^3}\int{d^2k_{\perp}dk_3dk_0}\left[J_+\left(\gamma^0k_0-\gamma^3k_3\right)
-2m_f\right] \\ && \nonumber\times\left[\frac{\delta
\left(k_{\parallel}^2-m_f^2\right)n_F(k_0)}{(p_\parallel-k_\parallel)^2-k_\perp^2}+\frac{\delta\left(\left(p_\parallel-k_\parallel\right)^2
-k_\perp^2\right)\left[-n_B(p_0-k_0)\right]}{k_\parallel^2-m^2_f}\right] 
\\ &\equiv& \Sigma_{n_F}(p_{\parallel})+\Sigma_{n_B}(p_{\parallel})
~.\end{eqnarray}
The part containing only the quark distribution function is rewritten as
\begin{eqnarray}
\Sigma_{n_F}(p_{\parallel}) = \frac{8g^{2}}{3(2\pi)^3}
\int{d^2k_{\perp}dk_3dk_0}\left[J_+\left(\gamma^0k_0-\gamma^3k_3\right)
-2m_f\right]\frac{\delta\left(k_0^2-\omega^2_k\right)n_F(k_0)}
{(p_\parallel-k_\parallel)^2-k_\perp^2}
~,\end{eqnarray}
where $\omega^2_k=k^2_3+m^2_f$. Using the property of 
Dirac delta function and considering the highest order 
of $p_\parallel^2$ as $|q_fB|$ in the strong magnetic 
field limit ($eB \gg T^2$) \cite{Gusynin:PLB450'1999}, 
integration over $k_0$ is done. So, we are left with 
$k_3$ and $k_\perp$ integrations,
\begin{eqnarray}\label{Sigma pp.}
\Sigma_{n_F}(p_{\parallel}) = \frac{-8g^{2}}{3(2\pi)^3}
\int{dk_3}\frac{n_F\left(\omega_k\right)}{\omega_k}
\left(J_+\gamma^3k_3+2m_f\right)\int{d^2k_\perp}\frac{1}
{p_\parallel^2+m_f^2-k^2_\perp}
~,\end{eqnarray}
where the $k_3$-integration is performed \cite{Dolan:PRD9'1974} to obtain
\begin{eqnarray}\label{$k_3$ integration}
I^3 = 4m_f\left[-\frac{1}{2}\ln\left(\frac{m_f}{\pi{T}}\right)-\frac{1}{2}\gamma_E+\mathcal{O}\left(\frac{m^2_f}{T^2}\right)\right]
.\end{eqnarray}
In the above equation, $\gamma_E$ represents the Euler-Mascheroni 
constant. The temperatures achieved in the heavy ion collisions are 
far larger than the masses of the light flavors ($u$ and $d$) 
{\em i.e.} $m_f\ll{T}$, so after dropping the term 
``$\mathcal{O}\left({m^2_f}/{T^2}\right)$", the 
eq. (\ref{$k_3$ integration}) becomes
\begin{eqnarray}
I^3=-2m_f\left[\ln\left(\frac{m_f}{\pi{T}}\right)+\gamma_E\right]
.\end{eqnarray}

Using the LLL approximation, we have done the 
$k_\perp$-integration in eq. (\ref{Sigma pp.}) and the 
real part is obtained as
\begin{eqnarray}
I^\perp &=& \pi\int_0^{|q_fB|} {dk_\perp^2} \frac{1}
{p_\parallel^2+m_f^2-k^2_\perp}\nonumber\\
&=&-\pi\ln\left(\frac{|q_fB|}{p_\parallel^2+m_f^2}\right)
.\end{eqnarray}
After substituting the values of $I^3$ and $I^\perp$, the part 
containing the quark distribution function (\ref{Sigma pp.}) becomes
\begin{eqnarray}\label{QSE in medium}
\Sigma_{n_F}(p_{\parallel}) &=& -\frac{2g^{2}m_f}{3\pi^2}
\ln\left(\frac{|q_fB|}{p_\parallel^2+m_f^2}\right)
\left[\ln\left(\frac{m_f}{\pi{T}}\right)+\gamma_E\right].
\end{eqnarray}

Now the medium part involving the gluon distribution function in 
eq. (\ref{S.D.F.P}) is rewritten as
\begin{eqnarray}\label{G.S.E. with G.D.F.}
\Sigma_{n_B}(p_{\parallel}) &=& -\frac{8g^{2}}{3(2\pi)^3}
\int {d^2k_{\perp}dk_3dk_0}\left[J_+\left(\gamma^0k_0-\gamma^3k_3\right)
-2m_f\right] \nonumber \\ && \times \frac{\delta\left(\left(p_\parallel-k_\parallel\right)^2
-k_\perp^2\right) n_B(p_0-k_0)}{k^2_0-\omega^2_k}
~.\end{eqnarray}
To solve the above equation we first find the roots of the following equation, 
\begin{eqnarray}
{(p_\parallel-k_\parallel)}^2
-k_\perp^2 = 0
~,\end{eqnarray}
then for small values of $k_\perp$ the above Dirac delta function 
is approximated as,
\begin{eqnarray}
\delta\left(\left(p_\parallel-k_\parallel\right)^2
-k_\perp^2\right)\approx
\frac{\left[\delta\{k_0-\left(p_0+p_3-k_3\right)\}
+\delta\{k_0-\left(p_0-p_3+k_3\right)\}\right]}{2\left(p_3-k_3\right)}
~.\end{eqnarray}
Using the above Dirac delta function, after performing the 
$k_0$, $k_\perp$ and $k_3$ integrations, we get the following 
form of the self-energy (\ref{G.S.E. with G.D.F.}),
\begin{eqnarray}
\nonumber\Sigma_{n_B}(p_{\parallel}) &=& -\frac{2|q_fB|g^{2}}{3(2\pi)^2}
\int^{+\infty}_{-\infty}{dk_3}~\frac{n_B(p_3-k_3)}{\left(p_3-k_3\right)}
\left[\frac{J_+\gamma^0p_0-2m_f}{\left(p_0+p_3-k_3\right)^2-\omega^2_k}\right. \\ && \left.\nonumber+\frac{J_+\gamma^0\left(p_3-k_3\right)}{\left(p_0+p_3-k_3\right)^2
-\omega^2_k}-\frac{J_+\gamma^3k_3}{\left(p_0+p_3-k_3\right)^2-\omega^2_k}+\frac{J_+\gamma^0p_0-2m_f}{\left(p_0-p_3+k_3\right)^2
-\omega^2_k}\right. \\ && \left.-\frac{J_+\gamma^0\left(p_3-k_3\right)}{\left(p_0-p_3+k_3\right)^2-\omega^2_k}-\frac{J_+\gamma^3k_3}{\left(p_0-p_3+k_3\right)^2
-\omega^2_k}\right] \nonumber\\
&=&-\frac{2|q_fB|g^{2}}{3(2\pi)^2} \sum_{i=1}^6 L^i
~,\end{eqnarray}
where $L^i$'s are the imaginary quantities with the following values,
\begin{eqnarray}
&&\nonumber{L^1}=\frac{i\pi\left(J_+\gamma^0{p_0}-2m_f\right)}{2\left(p_0+p_3\right)}
\left[\frac{\beta(a-p_3)-2}{2\beta(p_3-a)^2}+\frac{n_B(a-p_3)}{a-p_3}\right]~, \\
&&\nonumber{L^2}=\frac{-i\pi{J_+}\gamma^0}{2\left(p_0+p_3\right)}
\left[\frac{1}{\beta(p_3-a)}+n_B(a-p_3)\right]~, \\
&&\nonumber{L^3}=\frac{-i\pi{J_+}\gamma^3}{2\left(p_0+p_3\right)}
\left[\frac{-2a-\beta{p_3}(p_3-a)}{2\beta(p_3-a)^2}+\frac{an_B(a-p_3)}{a-p_3}\right]~, \\
&&\nonumber{L^4}=\frac{-i\pi\left(J_+\gamma^0{p_0}-2m_f\right)}{2\left(p_0-p_3\right)}
\left[\frac{\beta(b-p_3)-2}{2\beta(p_3-b)^2}+\frac{n_B(b-p_3)}{b-p_3}\right]~, \\
&&\nonumber{L^5}=\frac{-i\pi{J_+}\gamma^0}{2\left(p_0-p_3\right)}
\left[\frac{1}{\beta(p_3-b)}+n_B(b-p_3)\right]~, \\
&&\nonumber{L^6}=\frac{i\pi{J_+}\gamma^3}{2\left(p_0-p_3\right)}
\left[\frac{-2b-\beta{p_3}(p_3-b)}{2\beta(p_3-b)^2}+\frac{bn_B(b-p_3)}{b-p_3}\right]
~,\end{eqnarray}
where $a$ and $b$ are given by
\begin{eqnarray}
&&\nonumber{a}=\frac{\left(p_0+p_3\right)^2-m^2_f}{2\left(p_0+p_3\right)}
~,~ {b}=\frac{\left(p_0-p_3\right)^2-m^2_f}{2\left(p_3-p_0\right)}
~.\end{eqnarray}
Thus the medium part involving gluon distribution function 
($\Sigma_{n_B}(p_{\parallel})$) is excluded from the real part 
of the quark self-energy.

In eq. (\ref{Total sigma}), the medium part, containing the 
product of quark and gluon distribution functions,
\begin{eqnarray}
\nonumber\Sigma_{n^2}(p_{\parallel}) &=& \frac{-8g^{2}i}{3(2\pi)^2}
\int{d^2k_{\perp}{d^2k_\parallel}}\left[n_F(k_0)n_B(p_0-k_0)\right]\left[J_+\left(\gamma^0k_0-\gamma^3k_3\right)
-2m_f\right] 
\\ && \times\left[\delta(k_{\parallel}^2-m_f^2)
\delta((p_\parallel-k_\parallel)^2-k_\perp^2)\right]
\end{eqnarray}
is imaginary, hence it is not included in the real part of the 
quark self-energy.

Now the total real part of the one-loop quark self-energy in a strong 
magnetic field is obtained by adding the vacuum (\ref{QSE in vacuum}) 
and the medium (\ref{QSE in medium}) parts,
\begin{eqnarray}\label{Q.S.E.}
\nonumber\Sigma(p_{\parallel}) &=& \frac{{J_+}\left(\gamma^\parallel
\cdot{p}_\parallel\right)g^{2}}{12\pi^2}\left[-1
-\frac{|q_fB|}{m^2_f}\left\lbrace\ln\left(\frac{|q_fB|}{m^2_f}\right)
-1\right\rbrace\right] \\ && \nonumber+\frac{g^{2}m_f}{3\pi^2}\left[1
+\frac{{J_+}|q_fB|}{2m_f^2}\left\lbrace\ln\left(\frac{|q_fB|}{m^2_f}\right)
-1\right\rbrace\right] \\ && -\frac{2g^{2}m_f}{3\pi^2}
\ln\left(\frac{|q_fB|}{p_\parallel^2+m_f^2}\right)
\left[\ln\left(\frac{m_f}{\pi{T}}\right)+\gamma_E\right]
.\end{eqnarray}

\renewcommand{\theequation}{B.\arabic{equation}}
\section{Free energy due to the quark contribution}
The free energy due to quark contribution is written as
\begin{eqnarray}\label{F.E.Q.}
\mathcal{F}_q &=& -N_cN_f\int\frac{d^2p_\perp}{(2\pi)^2}\int\frac{d^2p_\parallel}
{(2\pi)^2}\ln\left[\det\left(\gamma^\parallel
\cdot{p}_\parallel-m_f-\Sigma(p_\parallel)\right)\right]
.\end{eqnarray}
In the presence of a strong magnetic field in the 
$z$ direction, the momentum splits into the 
components transverse and longitudinal to the 
magnetic field, so the momentum integration also 
gets factorized into transverse and longitudinal 
parts and in SMF limit, the integrand becomes a 
function of the longitudinal momentum only. To solve 
eq. (\ref{F.E.Q.}), we first write the quark self-energy 
(\ref{Q.S.E.}) in a simplified version as
\begin{eqnarray}
\Sigma(p_\parallel)&=&{J_+}\left(\gamma^\parallel
\cdot{p}_\parallel\right)\eta+{J_+}\Gamma+\xi+\Upsilon, ~{\rm{with}} \\
\eta&=&-\frac{g^{2}}{12\pi^2}\left[1
+\frac{|q_fB|}{m^2_f}\left\lbrace\ln\left(\frac{|q_fB|}
{m^2_f}\right)-1\right\rbrace\right]\label{C}, \\
\Gamma&=&\frac{g^{2}}{6\pi^2}\frac{|q_fB|}{m_f}
\left\lbrace\ln\left(\frac{|q_fB|}{m^2_f}\right)-1\right\rbrace\label{D}, \\ 
\xi&=&\frac{g^{2}m_f}{3\pi^2}\label{E}, \\ 
\Upsilon&=&-\frac{2g^{2}m_f}{3\pi^2}\ln\left(\frac{|q_fB|}{p_\parallel^2+m_f^2}\right)\left[\ln\left(\frac{m_f}{\pi{T}}\right)+\gamma_E\right]\label{F}.
\end{eqnarray}
The value of the determinant in eq. (\ref{F.E.Q.}) is given by
\begin{eqnarray}
\det\left[\gamma^\parallel\cdot{p}_\parallel-m_f-\Sigma(p_\parallel)\right] 
&=& \left[p^2_\parallel\left(1-2\eta\right)^2
-\left(m_f+2\Gamma+\xi+\Upsilon\right)^2\right] \nonumber\\ 
&& \times\left[p^2_\parallel-\left(m_f+\xi+\Upsilon\right)^2\right]
.\end{eqnarray}
After substituting the determinant into eq. (\ref{F.E.Q.}), we get
\begin{eqnarray}\label{F.E.Q.C.}
\nonumber\mathcal{F}_q &=& -N_cN_f
\int\frac{d^2p_\perp}{(2\pi)^2}\int\frac{d^2p_\parallel}
{(2\pi)^2}\ln\left[\Bigg\lbrace{p^2_\parallel}\left(1-2\eta\right)^2
-\left(m_f+2\Gamma+\xi+\Upsilon\right)^2\Bigg\rbrace\right. \\ && \left.\nonumber\hspace{5.49 cm}\times\Bigg\lbrace{p^2_\parallel}
-\left(m_f+\xi+\Upsilon\right)^2\Bigg\rbrace\right] 
\\ &\equiv& -\frac{N_cN_f|q_fB|}{2\pi}\left(I_1+I_2\right)
,\end{eqnarray}
where the integral $I_1$ is written as
\begin{eqnarray}
I_1 &=& \int\frac{dp_0}{2\pi}\int\frac{dp_3}
{2\pi}\ln\left(p^2_0-p^2_3\right)
~.\end{eqnarray}
Using the imaginary-time formalism at finite temperature, the 
$I_1$ integral can be solved, where the continuous energy 
integrals are discretized into Matsubara frequency 
($\omega_n$) sums, {\em i.e.} 
$\int\frac{dp_0}{2\pi}\rightarrow{T}\sum_n$. For quarks, 
$\omega_n=(2n+1)\pi{T}$, where $n=0,1,2,\cdots$. After solving, 
we got the expression for $I_1$ as
\begin{eqnarray}
{I_1} &=& \frac{\pi{T^2}}{6}
~.\end{eqnarray}
The $I_2$ integral is given by
\begin{eqnarray}
\nonumber{I_2} &=& \int\frac{d^2p_\parallel}{2(2\pi)^2}\ln\left[\left\lbrace\left(1-2\eta\right)^2
-\frac{1}{p^2_\parallel}\left(m_f+2\Gamma+\xi+\Upsilon\right)^2\right\rbrace\left\lbrace 1
-\frac{1}{p^2_\parallel}\left(m_f+\xi+\Upsilon\right)^2\right\rbrace\right] 
\\ &=& \nonumber\int^{|q_fB|}_0\frac{dp^2_\parallel}{8\pi}\ln\left[\left\lbrace\left(1-2\eta\right)^2
-\frac{1}{p^2_\parallel}\left[m_f+2\Gamma+\xi+\lambda\ln\left(\frac{|q_fB|}{p_\parallel^2+m_f^2}\right)\right]^2\right\rbrace\right. \\ && \left.\hspace{2.49 cm}\times\left\lbrace 1
-\frac{1}{p^2_\parallel}\left[m_f+\xi+\lambda\ln\left(\frac{|q_fB|}{p_\parallel^2+m_f^2}\right)\right]^2\right\rbrace\right]
,\end{eqnarray}
where we have reexpressed the momentum-dependent term, $\Upsilon$ as
\begin{eqnarray}
\Upsilon &=& \lambda\ln\left(\frac{|q_fB|}{p_\parallel^2+m_f^2}\right),~
{\rm{with}}\\
\lambda &=& -\frac{2g^{2}m_f}{3\pi^2}\left[\ln\left(\frac{m_f}{\pi{T}}\right)
+\gamma_E\right]\label{W}
.\end{eqnarray}
After plugging the integrals $I_1$ and $I_2$ in eq. (\ref{F.E.Q.C.}), the 
free energy due to the quark contribution in a strong magnetic field is expressed as,
\begin{eqnarray}
\nonumber\mathcal{F}_q &=& -\frac{N_cN_f|q_fB|}{4}
\left[\frac{T^2}{3}+\int^{|q_fB|}_0\frac{dp^2_\parallel}{4\pi^2}\ln\left[\Bigg\lbrace\left(1-2\eta\right)^2\right.\right. \\ && \left.\left.
\nonumber-\frac{1}{p^2_\parallel}\left[m_f+2\Gamma+\xi+\lambda\ln\left(\frac{|q_fB|}{p_\parallel^2+m_f^2}\right)\right]^2\Bigg\rbrace\right.\right. \\ && \left.\left.\times\left\lbrace 1
-\frac{1}{p^2_\parallel}\left[m_f+\xi+\lambda\ln\left(\frac{|q_fB|}{p_\parallel^2+m_f^2}\right)\right]^2\right\rbrace\right]\right]
.\end{eqnarray}

\renewcommand{\theequation}{C.\arabic{equation}}
\section{Integrals}
\begin{eqnarray}\label{I.1}
\nonumber\int\frac{d^3\mathbf{p}}{(2\pi)^3}\ln\left(\mathbf{p}^2+m^2_{q,D}\right) 
&=& \mu^{2\epsilon}\int\frac{d^{3-2\epsilon}\mathbf{p}}{(2\pi)^{3-2\epsilon}}\ln\left(\mathbf{p}^2+m^2_{q,D}\right) \\ &=& 
\nonumber\frac{\mu^{2\epsilon}}{2^{2-2\epsilon}\pi^{\frac{3-2\epsilon}{2}}}
\frac{1}{\Gamma\left(\frac{3-2\epsilon}{2}\right)}
\int^\infty_0d\mathbf{p}~\mathbf{p}^{2-2\epsilon}\ln\left(\mathbf{p}^2
+m^2_{q,D}\right) \\ &=& -\frac{m^3_{q,D}}{3}\left(\frac{\mu}{2m_{q,D}}\right)^{2\epsilon}
\left[\frac{1}{2\pi}+\frac{4\epsilon}{3}\right]
,\end{eqnarray}
where $\mu$ is an arbitrary mass parameter. Setting $\epsilon=0$, we get
\begin{eqnarray}
\int\frac{d^3\mathbf{p}}{(2\pi)^3}\ln\left(\mathbf{p}^2+m^2_{q,D}\right) &=& -\frac{m^3_{q,D}}{6\pi}
~.\end{eqnarray}
\begin{eqnarray}\label{I.2}
\nonumber\int\frac{d^4P}{(2\pi)^4}~\frac{\Pi_\parallel}{\mathbf{p}^2} 
&=& -\sum_f\int\frac{d^3\mathbf{p}}{(2\pi)^3}\int\frac{dp_0}{2\pi}~\frac{g^2|q_fB|p_3^2}{8\pi^2p_\parallel^2\mathbf{p}^2} \\ &=& -\frac{g^2}{8\pi^2}\sum_f|q_fB|\int\frac{d^3\mathbf{p}}{(2\pi)^3}T\sum_n\frac{1}{p_0^2-p_3^2}~\frac{p_3^2}{\mathbf{p}^2}
~,\end{eqnarray}
where the frequency sum is given by
\begin{eqnarray}\label{I.3}
T\sum_n\frac{1}{p_0^2-p_3^2}=-\frac{1}{2p_3}\left\lbrace1+2n_B(p_3)\right\rbrace
.\end{eqnarray}
Now eq. (\ref{I.2}) becomes
\begin{eqnarray}\label{I.4}
\int\frac{d^4P}{(2\pi)^4}~\frac{\Pi_\parallel}{\mathbf{p}^2} 
&=& \frac{g^2}{16\pi^2}\sum_f|q_fB|\int\frac{d^3\mathbf{p}}{(2\pi)^3}~\frac{p_3}{\mathbf{p}^2}\left\lbrace1+2n_B(p_3)\right\rbrace
.\end{eqnarray}
For $p_3<T$, we solve the above integral and the simplified value is given by
\begin{eqnarray}\label{I.5}
\int\frac{d^4P}{(2\pi)^4}~\frac{\Pi_\parallel}{\mathbf{p}^2} 
&=& \sum_f\frac{g^2T|q_fB|\Omega}{4(2\pi)^3}
~,\end{eqnarray}
where $\Omega$ is the ultraviolet cutoff parameter.
\begin{eqnarray}\label{I.6}
\nonumber\int\frac{d^4P}{(2\pi)^4}~\frac{\Pi_\parallel^2}{2\mathbf{p}^4} 
&=& \frac{g^4}{128\pi^4}\sum_f|q_fB|^2
\int\frac{d^3\mathbf{p}}{(2\pi)^3}\int\frac{dp_0}{2\pi}~\frac{p_3^4}
{\mathbf{p}^4p_\parallel^4} \\ &=& \frac{g^4}{128\pi^4}\sum_f|q_fB|^2
\int\frac{d^3\mathbf{p}}{(2\pi)^3}~\frac{p_3^4}{\mathbf{p}^4}T\sum_n\frac{1}
{\left(p_0^2-p_3^2\right)^2}
~.\end{eqnarray}
The frequency sum in the above equation is performed as follows.
\begin{eqnarray}\label{I.7}
\nonumber{T\sum_n\frac{1}{\left(p_0^2-p_3^2\right)^2}} 
&=& T\sum_n\frac{1}{2p_3}\frac{\partial}{\partial{p_3}}
\left(\frac{1}{p_0^2-p_3^2}\right) \\ &=& \frac{1}{4p_3^3}\left\lbrace1+2n_B(p_3)\right\rbrace+\frac{1}{2Tp_3^2}\left\lbrace1+n_B(p_3)\right\rbrace{n_B(p_3)}
~.\end{eqnarray}
Substituting the frequency sum in eq. (\ref{I.6}), we 
obtain the terms involving infrared divergence. So, in 
solving the integration, we use cutoff regularization 
method to discard the divergence and thus, eq. (\ref{I.6}) 
is simplified into
\begin{eqnarray}\label{I.8}
\int\frac{d^4P}{(2\pi)^4}~\frac{\Pi_\parallel^2}{2\mathbf{p}^4} 
&=& \sum_f\frac{g^4T|q_fB|^2}{128\pi^2(2\pi)^3\zeta}
~,\end{eqnarray}
where $\zeta$ represents the infrared cutoff parameter.

\end{appendices}

\end{document}